\documentclass[
reprint,
superscriptaddress,
amsmath,amssymb,
aps
]{revtex4-2}
\usepackage{flushend}

\usepackage{graphicx}
\usepackage{dcolumn}
\usepackage{bm}
\usepackage{hyperref}
\usepackage[mathlines]{lineno}
\usepackage{xcolor}

\begin{document}

\title{Higher-Order Competition as a Minimal Mechanism for Spatial Pattern Diversity in Population Dynamics}

\author{David Pinto-Ramos}
 \affiliation{Center for Advanced Systems Understanding (CASUS), Helmholtz-Zentrum Dresden-Rossendorf (HZDR), D-02826 Görlitz, Germany}
\author{Anudeep Surendran}%
\affiliation{Center for Advanced Systems Understanding (CASUS), Helmholtz-Zentrum Dresden-Rossendorf (HZDR), D-02826 Görlitz, Germany}%

\author{Ricardo Martinez-Garcia}
\email{r.martinez-garcia@hzdr.de}
\affiliation{Center for Advanced Systems Understanding (CASUS), Helmholtz-Zentrum Dresden-Rossendorf (HZDR), D-02826 Görlitz, Germany}
\affiliation{ICTP South American Institute for Fundamental Research \& Instituto de F\'isica Te\'orica, Universidade Estadual Paulista - UNESP, Brazil.}
\affiliation{Department of Ecology, Institute of Biosciences, University of São Paulo, São Paulo, Brazil }

\date{\today}

\begin{abstract}

We show that negative feedback alone generates the diversity of self-organized shapes usually attributed to scale-dependent activation-inhibition. For a broad class of kernels, pairwise competition models only generate hexagonal spot arrays. Higher-order terms eliminate this restriction and promote stripes and gaps. Combining individual-based simulations and nonlinear analysis, we derive the pattern-selection thresholds and construct the full state diagram, including an unusual spots-stripes-spots sequence. Spot patterns are thus not a reliable indicator of proximity to a tipping point.
\end{abstract}

\maketitle

\textit{Introduction--}Spatial self-organization is ubiquitous in physical, chemical, and biological systems \cite{Lugiato1987, Cross1993,Bailles2022,Martinez-Garcia2022,Pringle2017,Rietkerk2008}. Periodic patterns form when space-dependent perturbations amplify specific modes of an unstable uniform state until nonlinearities, arising from microscopic interactions among system components, saturate this growth and select the resulting pattern's morphology \cite{Turing1952, Cross1993}. Understanding how such microscopic interactions control pattern selection and lead to the diversity of structures observed in nature remains a central problem in pattern formation theory and its applications \cite{Meron2019}. 

In several examples, such diversity arises from the interplay between short-range positive and long-range negative interactions, including activation–inhibition scale-dependent feedback in reaction–diffusion models and attraction–repulsion dynamics in many nonlinear diffusion equations \cite{Kondo2010, Liu2016}. Alternatively, some studies show that negative feedback alone is sufficient to generate pattern-forming instabilities \cite{martines2013vegetation, delfau2016pattern}. 

A paradigmatic model that allows for pattern formation via negative feedback alone is the nonlocal Fisher–Kolmogorov–Petrovskii–Piskunov (FKPP) equation, introduced to study propagation phenomena \cite{Gourley2000, Berestycki2009} and widely adopted in population dynamics \cite{Fuentes2003, lopezFluctuationsImpact2004, Hernandez-Garcia2004, piva2021interplay, Silvano2025}. In this model, competition arises from pairwise interactions alone, yielding a mortality term linear in the nonlocally averaged population density that can destabilize homogeneous distributions of organisms \cite{Fuentes2003, Pigolotti2007}. 

However, previous studies of the standard two-dimensional nonlocal FKPP equation reported only hexagonal arrays of population clusters. Natural populations exhibit much richer morphological diversity, including stripes, labyrinths, and gaps \cite{Rietkerk2008, Martinez-Garcia2022}. Recovering these shapes has required additional dynamical fields, nonlocal terms representing positive interactions, or nonlinearities with no clear individual-level origin \cite{VonHardenberg2001,borgogno2009mathematical,martines2013vegetation,tlidi2024non}. Each of these constructions departs from the minimal construction of the nonlocal FKPP and cannot inform what a competing population minimally needs to form patterns other than spots.

Whether these hexagons are an intrinsic feature of pairwise competition or are simply the only patterns found so far remains unknown. Pattern shape is selected by the nonlinear terms of the density-field equation, and a nonlinear selection theory for the two-dimensional nonlocal FKPP has not been developed (but see \cite{lopezFluctuationsImpact2004} for partial results). Linear stability analysis fixes the pattern's characteristic wavelength, but provides no information about its symmetries. Consequently, two questions remain unanswered: Are hexagonal arrays of spots the only pattern resulting from two-dimensional pairwise competition? If so, what is the minimal demographic feedback that breaks this limitation?

We address both questions analytically. First, we show that pairwise competition restricts stable patterns to hexagonal spot arrays. We then show that higher-order competition provides a minimal demographic feedback that generates the full diversity of spots, labyrinths, and gaps observed in nature. Such competition arises naturally when a locally depleted shared resource mediates the interaction \cite{Mendez2019,Jorge2023}. Combining stochastic spatial birth-death simulations, numerical solutions of the corresponding density-field equation, and an invariant-manifold reduction, we obtain the full diagram of pattern configurations, including analytical expressions for each pattern selection threshold. This diagram contains an unusual spots--stripes--spots sequence, which questions the use of spot patterns as indicators of proximity to ecological tipping points \cite{rietkerk2004self,franklinOrganizingPrinciplesVegetation2020,rietkerk2021evasion}.

\begin{figure*}[t]
	\centering
	\includegraphics[width=0.9\textwidth]{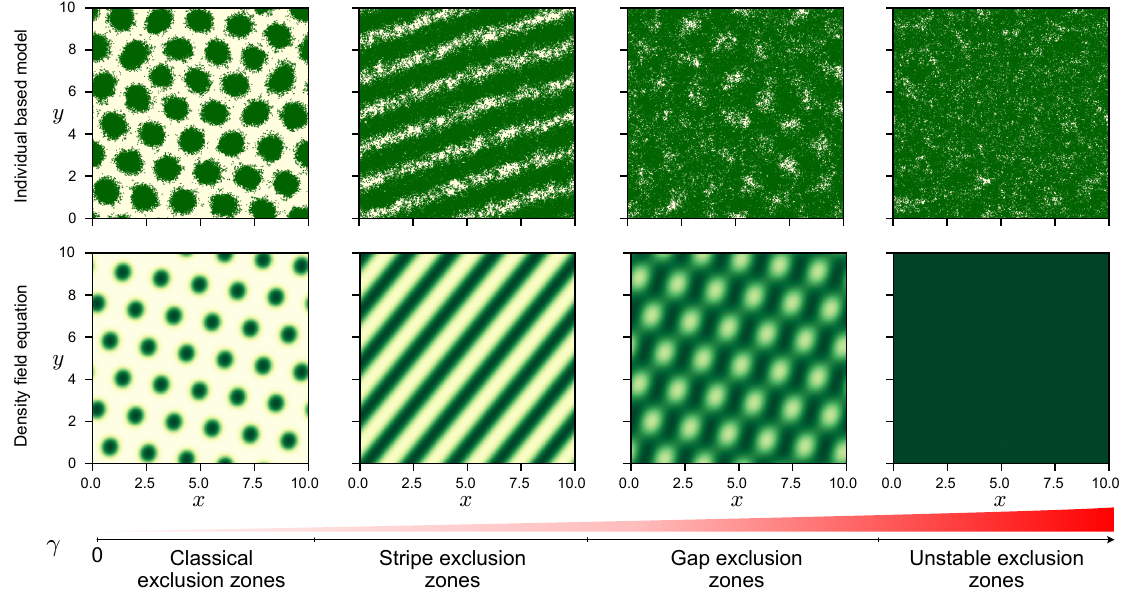}
	\caption{Higher-order competition generates a diversity of spatial patterns. Individual-based stochastic simulations (top row) and the corresponding density-field equation (bottom row) produce the same sequence of pattern shapes. As the strength of higher-order competition, $\gamma$, increases (left to right), the hexagonal symmetry of the spot pattern breaks and the system transitions to stripes and then to gaps. $\gamma$ increases from left to right: $3.27 \times 10^{-6}$, $2.35\times 10^{-5}$, $3.14\times10^{-5}$, $3.93\times10^{-5}$. Other parameters are held constant: $b=1920$, $d_0=192$, $\eta=1$, $r_c=1.25$, $s=0.05$, $r_{h.o.}=0.0625$. }
	\label{F1}
\end{figure*}

\textit{Discrete model--} Consider a variable number $N$ of individuals at positions $\{\mathbf{r}_i\}$ within a rectangular domain of area $L^2$. The population size changes following a spatially explicit birth-death dynamics, which we model as a one-step continuous Markov process with two processes. i) \textit{Reproduction and dispersal--} an individual at position $\mathbf{r}$ reproduces, with fixed birth rate $b$, and the offspring is placed randomly in a position $\mathbf{r'}$ drawn from a probability density $\phi_D(\mathbf{r}-\mathbf{r'})$, representing a dispersal kernel. ii) \textit{Death and competition--} an individual at position $\mathbf{r}$ can die with a total death rate $d(\mathbf{r}) = d_0 + d_c(\mathbf{r})$, where $d_0$ is a constant baseline mortality and $d_c(\mathbf{r})$ is a density-dependent contribution due to competition. We consider
\begin{equation}
d_c(\mathbf{r}) = \eta \sum_{i}^{N} \phi_{p}(\mathbf{r}-\mathbf{r}_i) + \gamma \sum_{i}^{N}\sum_{j}^{N}\psi_{\mathrm{\tiny{H.O.}}}(\mathbf{r}, \mathbf{r}_i, \mathbf{r}_j),
\end{equation}
where $\phi_p$ is the pairwise competition kernel of strength $\eta$, which leads to the classical nonlocal FKPP equation, and $\gamma$ is the intensity of the three-body higher-order competition, mediated by the three-body kernel $\psi_{\mathrm{\tiny{H.O.}}}$. A non-zero $\gamma$ thus implies that the competition felt by an individual is not just the superposition of its pairwise interactions with the rest of the population. The kernels are assumed to be normalized to unity so the interaction strength scales with density and not with abundance. 

We first simulated this individual-based stochastic dynamics (Fig.\,\ref{F1}; top row) using a Gaussian dispersal kernel $\phi_D$ with zero mean and variance $s^2$, a top-hat pairwise competition kernel $\phi_p$ with radius $r_c$, and a higher-order competition kernel of the form $\psi_{\mathrm{\tiny{H.O.}}}= \phi_{h.o.} (\mathbf{r}-\mathbf{r}_i)\phi_{h.o.}(\mathbf{r}-\mathbf{r}_j)$ where $\phi_{h.o.}$ is a top-hat function of radius $r_{h.o.}$. For purely pairwise competition, $\gamma=0$, we recover the usual behavior of the nonlocal FKPP equation \cite{surendran2025spatial}, which, for sufficiently high birth rates, exhibits hexagonally arranged clusters of organisms separated by unpopulated exclusion zones of high mortality. As $\gamma$ increases, however, the clusters first stretch and form a stripe pattern, which becomes a gap pattern when $\gamma$ is further increased. Higher-order interactions therefore suppress the exclusion zones, which are usually seen as the mechanism behind patterns in the classical nonlocal FKPP equation, where only hexagonal spots are found \cite{martinez2023integrating}.

\textit{Coarse-grained model--} To understand how higher-order competition controls pattern formation and shape, we derive the coarse-grained density equation corresponding to the individual-based stochastic dynamics. We first discretize the spatial domain into a regular grid of $M^2$ cells with lattice spacing $dx = L/M$. Each cell, labeled by $(i,j)$, contains $N_{i,j}$ individuals, so the state of the population is fully specified by the vector $\mathbf{\Omega} =\{N_{1,1}, N_{1,2}, \dots, N_{M,M}\}$. This state evolves according to the probabilistic birth-death rules described above, so $P(\mathbf{\Omega}, t)$ denotes the probability of observing each configuration at time $t$.

The dynamics of this probability mass function $P(\mathbf{\Omega},t)$ is given by a Master equation of the form
\begin{equation}
\label{eq:master}
\frac{\partial P}{\partial t} \!=\! \sum_{\mathbf{\Omega'}} \Big[W(\mathbf{\Omega'}\!\rightarrow \!\mathbf{\Omega}) P(\mathbf{\Omega'}) \!-\! W( \mathbf{\Omega}\!\rightarrow \!\mathbf{\Omega'}) P(\mathbf{\Omega})\Big],
\end{equation}
where $W(\mathbf{\Omega}\rightarrow\mathbf{\Omega}')$ is the transition rate from configuration $\mathbf{\Omega}$ to $\mathbf{\Omega}'$. Because the underlying birth-death process is a one-step process, every event changes the occupation of a single cell by exactly $\pm1$. We therefore denote by $\mathbf{\Omega}_{k,l}^{\pm}$ the configuration obtained from $\mathbf{\Omega}$ by adding or removing one individual in cell $(k,l)$, and abbreviate $W_{k,l}^{\pm} \equiv W(\mathbf{\Omega}\rightarrow\mathbf{\Omega}_{k,l}^{\pm})$. These rates are
\begin{align}
W_{k,l}^{+} &= \sum_{i,j} b\, N_{i,j}\, \phi_D(\mathbf{r}_{i,j}-\mathbf{r}_{k,l})\, dx^2, \\
W_{k,l}^{-} &= N_{k,l} \bigg[ d_0 + \eta \sum_{i,j} N_{i,j}\, \phi_p(\mathbf{r}_{i,j}-\mathbf{r}_{k,l}) \nonumber \\
&\quad + \gamma\sum_{i,j}\sum_{m,n} N_{i,j} N_{m,n}\,
\psi_{\mathrm{H.O.}}(\mathbf{r}_{k,l}, \mathbf{r}_{i,j}, \mathbf{r}_{m,n}) \bigg]
\end{align}
and they define the stochastic process completely. The resulting Master equation, however, is analytically intractable. We therefore derive a macroscopic description in terms of a continuous mean density field $n(\mathbf{r},t)$, which we construct in three steps. First, we take the first moment of Eq.\,\eqref{eq:master}, multiplying by $N_{k,l}$ and summing over all possible configurations, $\langle N_{k,l} \rangle = \sum_{\mathbf{\Omega}} N_{k,l}\, P(\mathbf{\Omega},t)$, to obtain an equation for $\partial_t \langle N_{k,l} \rangle$ in terms of the rates $W_{k,l}^{\pm}$. Second, because the death rate is nonlinear in the occupation numbers, the dynamics of the first moment is coupled to the second and third moments, so the equation for $\langle N_{k,l} \rangle$ is not closed. We apply a mean-field closure, which neglects spatial correlations by factorizing higher-order moments into products of means $\langle N_{i,j} N_{k,l} \rangle = \langle N_{i,j} \rangle \langle N_{k,l} \rangle$ and $\langle N_{i,j} N_{k,l} N_{m,n} \rangle = \langle N_{i,j} \rangle \langle N_{k,l} \rangle \langle N_{m,n} \rangle$. Finally, writing $n(\mathbf{r}_{i,j}) = \langle N_{i,j} \rangle / dx^2$ and taking $dx$ sufficiently small at fixed density (see the \textit{Supplementary Materials} \ref{sec:SM-meanfield} for further details), we obtain
\begin{equation}
\begin{split}
\frac{\partial n}{\partial t} &= b\!\int\! \phi_D(\mathbf{r}-\mathbf{r}')\, n(\mathbf{r}')\, \mathrm{d}\mathbf{r}' \\
&\quad - n(\mathbf{r}) \bigg( d_0 + \eta\int\! \phi_p(\mathbf{r}-\mathbf{r}')\, n(\mathbf{r}')\, \mathrm{d}\mathbf{r}' \\
&\quad + \gamma\!\iint\! \psi_{\mathrm{H.O.}}(\mathbf{r},\mathbf{r}',\mathbf{r}'')\, n(\mathbf{r}')n(\mathbf{r}'')\, \mathrm{d}\mathbf{r}'\mathrm{d}\mathbf{r}'' \bigg),
\end{split}
\label{ho_nl_fkpp}
\end{equation}
\noindent where $\mathrm{d}\mathbf{r}=\mathrm{d}x\mathrm{d}y$ is the infinitesimal area element. We refer to this density-field equation (DFE) as the \textit{higher-order nonlocal FKPP equation}, which reduces to the standard nonlocal FKPP equation \cite{Achleitner2015,piva2021interplay,surendran2025spatial} when higher-order interactions are neglected, $\gamma = 0$, and dispersal is short-ranged. For short-range dispersal, the kernel admits a gradient expansion $b\!\int\! \phi_D(\mathbf{r}-\mathbf{r}')\, n(\mathbf{r}')\, \mathrm{d}\mathbf{r}' \approx b\, n(\mathbf{r}) + D \nabla^2 n(\mathbf{r})$, where $4D = b\int |\mathbf{r}|^2 \phi_D(\mathbf{r})\, \mathrm{d}\mathbf{r}$. 

Equation \eqref{ho_nl_fkpp} admits the rescaling $n= \tilde{n}b/\eta$ and $t=\tilde{t}/b$, which absorbs the quadratic coefficient. We therefore set $\eta=1$ in what follows. Absorbing the highest-order non-linearity would be the most standard choice, but we keep the cubic coefficient free so we can vary the strength of higher-order competition explicitly. This choice leaves two dimensionless parameters free: the ratio of baseline mortality to birth rate, $d_0/b$, and $\gamma b/\eta^2$, which sets the strength of higher-order relative to pairwise competition.

Numerical solutions of this scaled DFE\,\eqref{ho_nl_fkpp} reproduce the same sequence of patterns as the individual-based stochastic simulations (IBM) (Fig.\,\ref{F1}, bottom row; see \ref{sec:SM-num} for details of the implementation). We compare them quantitatively using the structure tensor $S(\mathbf{r}) = \langle \nabla n \otimes \nabla n \rangle_{\text{window}}$, where the average runs over a window wider than the pattern wavelength. The eigenvalues of this tensor, $l_{1,2}(\mathbf{r})$, are the mean squared density gradient along the two
principal directions, so their normalized difference, $\mathcal{A}_l(\mathbf{r})=\rvert l_1 - l_2\rvert / (l_1 + l_2)$, vanishes for locally isotropic patterns (e.g., spots and gaps) and approaches unity for stripes. We define
the anisotropy index $\mathcal{A}$ as the median of $\mathcal{A}_l$ over the
domain and across realizations (see the \textit{Supplementary Materials} \ref{sec:SM-anisotropy} for implementation details).

\begin{figure}[t]
	\centering
	\includegraphics[width=0.9\columnwidth]{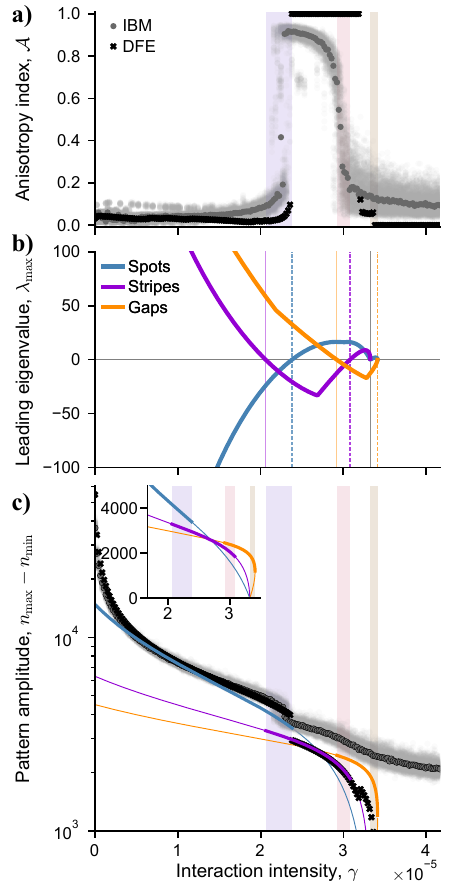}
	\caption{Nonlinear analysis of the birth-death dynamics. a) Anisotropy index. Gray dots show 10 IBM realizations, each with 20 snapshots, and dark dots show their median. Black crosses show the DFE median over 32 initial conditions. b) Leading eigenvalues of each pattern solution. Vertical continuous lines mark where these eigenvalues become negative as $\gamma$ increases (violet, orange), and the Turing instability (black); vertical dashed lines indicate $\gamma$ levels at which they become positive (blue, violet), and the saddle node of the hexagonal symmetry branch (orange). c) Pattern amplitude from the same simulations as in a). Thick lines show the predictions of Eq.\,\eqref{eq_hexagons}, and thin lines show the unstable branches. The inset shows the region around $\gamma_c$.}
	\label{F2}
\end{figure}

This anisotropy index captures the changes in pattern morphology as a function of $\gamma$ (Fig.\,\ref{F2}a). Past a threshold in $\gamma$, anisotropy increases abruptly from $0$ to $1$ as isotropic spots rearrange into stripes. Further increasing $\gamma$, $\mathcal{A}$ decreases again, finally saturating at a constant value. For the DFE, the high $\gamma$ limit is a perfectly uniform state for which both eigenvalues vanish and the normalized difference is undefined. Therefore, we assign $\mathcal{A}=0$ by convention, which is the value expected for a state with no preferred direction. IBM simulations, however, tend to a non-zero value because demographic fluctuations introduce residual heterogeneity in the spatial pattern.

\textit{Nonlinear analysis--}The agreement between DFE and IBM simulations justifies treating the deterministic equation as an analytically tractable approximation of the stochastic dynamics. We now use it to derive the pattern-selection thresholds introduced by higher-order competition. These thresholds follow from the equations governing the slow dynamics of the pattern amplitudes. The quadratic terms couple triplets of wavevectors that sum to zero, which for equal magnitudes forces relative angles of $\pm 2\pi/3$. We therefore restrict the expansion to the canonical resonant triad, $\mathbf{k}_1 + \mathbf{k}_2 + \mathbf{k}_3 = 0$, and write the density field as
\begin{equation}
n(\mathbf{r},t) = n_0 + \sum_{j=1}^3 A_j(t) e^{\mathbf{i} \mathbf{k}_j\cdot \mathbf{r}}+ \overline{A}_j(t) e^{-\mathbf{i} \mathbf{k}_j\cdot \mathbf{r}}+\text{h.o.t.},
\label{Eq:ansatz-nonlin}
\end{equation}
where $n_0(\gamma)=[-1+\sqrt{1+4\gamma(b-d_0)}]/(2\gamma)$ is the uniform solution of Eq.\,\eqref{ho_nl_fkpp}, $A_i(t)$ are complex amplitudes, $\mathbf{i}$ is the imaginary unit, $\overline{(\cdot)}$ denotes the complex conjugate, and h.o.t. stands for higher-order terms. All three wavevectors have the same magnitude $k_c$, which corresponds to the first wavenumber that becomes unstable as $\gamma$ decreases from the uniform density regime and crosses a critical value $\gamma_c$. 

Substituting Eq.\,\eqref{Eq:ansatz-nonlin} into Eq.\,\eqref{ho_nl_fkpp} and projecting on the resonant triad with slow dynamics (see the \textit{Supplementary Materials} \ref{sec:SM-ampl}), we obtain
\begin{equation}
\frac{d A_i}{d t} \!=\! \mu A_i - \kappa \overline{A}_j\overline{A}_k - A_i\!\Big[c|A_i|^2 \!+ \!\nu\left( |A_j|^2\!+\!|A_k|^2\right)\!\Big],
\label{eq_hexagons}
\end{equation}
with the remaining two equations obtained by cyclic permutation of the indices. The linear coefficient $\mu(\gamma)$ is the growth rate of the critical mode $k_c$, and $\kappa(\gamma)$, $c(\gamma)$, and $\nu(\gamma)$ set the quadratic and cubic saturation (see the \textit{Supplementary Materials} \ref{sec:SM-ampl} for their expressions in terms of the original parameters). 

Instead of evaluating these coefficients at the critical point $\gamma_c$, as is standard in weakly nonlinear analysis, we retain the full dependence on the bifurcation parameter $\gamma$. This allows us to track pattern selection far from pattern onset, where transitions between pattern shapes occur. Equation \eqref{eq_hexagons} admits two families of stationary solutions. Stripes
correspond to $|A_i| = \sqrt{\mu/c}$ with $A_j = A_k = 0$, and hexagons to
$|A_i| = |A_j| = |A_k| = \rho_{\pm}= [-\kappa \cos\Phi_0 \pm \sqrt{\kappa^2 + 4\mu(c+2\nu)}]/[2(c+2\nu)]$,
where the global phase $\Phi_0$ distinguishes spots ($\Phi_0 = 0$) from gaps
($\Phi_0 = \pi$) and is fixed by the sign of $\kappa$.

We determine the pattern selection thresholds by linearizing  Eq.\,\eqref{eq_hexagons} around the three canonical pattern solutions, yielding three eigenvalue problems. The solutions are: $\lambda_\mu=-2\mu$ and $\lambda_{\pm}=\mu(1-\nu/c)\pm\kappa\sqrt{\mu/c}$ for the stripe pattern; $\lambda_\Phi=3\kappa \rho_+ \cos\Phi_0$, $\lambda_{\text{sym}}=-2(c+2\nu)\rho_+^2-\kappa\rho_+\cos\Phi_0$, and $\lambda_{\text{asym}}=-2(c-\nu)\rho_+^2+2\kappa\rho_+\cos\Phi_0$ for spot ($\Phi_0=0$) and gap ($\Phi_0=\pi$) patterns. The leading eigenvalue of the linear problem around each pattern solution changes sign at values of $\gamma$ that delimit the region where stripes, gaps, and spots are observed in both the IBM and DFE simulations (Fig.\,\ref{F2}b). These transitions are also visible in the pattern amplitudes (Fig.\,\ref{F2}c). 

In the pairwise limit, $\gamma=0$, these formulas exclude both gaps and stripes analytically.
The sign of $\kappa$ determines if gaps or spots are selected and simultaneously controls the stability of stripes. For $\gamma=0$ we have $\kappa=2\hat{\phi}_p(k_c)$, and pattern formation in the traditional nonlocal FKPP requires $\hat{\phi}_p(k_c)<0$, so $\kappa(\gamma\to0)<0$. The phase eigenvalue of gap patterns is then non-negative and gaps are unstable or marginal at best \cite{lopezFluctuationsImpact2004}. For stripes, a simple set of sufficient conditions for their instability is $\hat{\phi}_D (qk_c)-2\hat{\phi}_D(k_c)+1<0$ for $q=\{2, \sqrt{3}\}$, $\hat{\phi}_p(k_c)>-1/6$, $\hat{\phi}_p(2k_c)>\hat{\phi}_p(k_c)/2$, and $\hat{\phi}_p(\sqrt{3}k_c)<-\hat{\phi}_p(k_c)$ for all admissible critical wavenumbers. These conditions hold for Gaussian dispersal kernels and their short-range diffusive approximation, for which $\hat{\phi}_D=1-Dk^2$, combined with a super-Gaussian competition kernel, $\phi_p \propto e^{-|\mathbf{x}/r_c|^p}$, which recovers the top-hat kernel for $p\to\infty$ limit. Because $\kappa<0$, the leading stripe eigenvalue is $\lambda_-$, and under those conditions it is always positive (see \textit{Supplementary Materials} for details). Hexagonally arranged spots are therefore the only stable member of the canonical family in the two-dimensional nonlocal FKPP. Patterns outside the resonant triad, such as square or rhombic lattices, are not covered by Eq.\,\eqref{eq_hexagons} and thus excluded from our analysis.

\begin{figure}[t!]
	\centering
	\includegraphics[width=0.9\columnwidth]{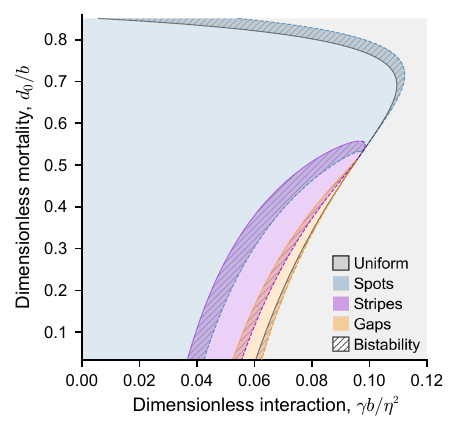}
	\caption{Diagram of states of Eq.\,\eqref{ho_nl_fkpp} in the plane of relative mortality $d_0/b$ and rescaled intensity of higher-order competition $\gamma b/\eta^2$. Shaded regions give the morphology predicted by the amplitude equations. Lines correspond to the selection thresholds depicted in Fig.\ref{F2}. The textured areas depict regions of bistability between morphologies adjacent in the diagram.}
	\label{F3}
\end{figure}

Higher-order competition eliminates this restriction. To see where each morphology is
selected, we construct the diagram of states as a function of the two dimensionless demographic parameters introduced in Eq.\,\eqref{ho_nl_fkpp}:
the relative mortality $d_0/b$ and the rescaled intensity of higher-order competition $\gamma b/\eta^2$. The other remaining parameters are the interaction ranges $s$, $r_c$, and $r_{\mathrm{h.o.}}$, which we keep fixed. Evaluating the eigenvalues above across the parameter space defined by $d_0/b$ and $\gamma b/\eta^2$, we can construct the diagram of states of the canonical pattern configurations (Fig.\,\ref{F3}). At low mortality, increasing $\gamma b/\eta^2$ drives the typical sequence of spots\,$\rightarrow$\,stripes\,$\rightarrow$\,gaps before the population becomes uniform, in agreement with the simulations of Fig.\,\ref{F1}. Near $d_0/b \simeq 0.55$, however, the stripe region closes, and the system re-enters the spotted state when $\gamma b/\eta^2$ increases, leading to an atypical sequence of spots\,$\rightarrow$\,stripes\,$\rightarrow$\,spots\,$\rightarrow$\,uniform. This re-entrance is driven by the nonlinear modulation that higher-order competition imposes on the amplitude-equation parameters, and we identify it because we retain their full $\gamma$ dependence instead of evaluating them at the pattern onset. 

\textit{Discussion--}Since Turing's work, an activation-inhibition scale-dependent feedback has been the prevailing explanation for the diversity of spatial patterns in natural populations \cite{Turing1952, Kondo2010,Rietkerk2008}. Negative feedback alone, interpreted as ecological competition, has been (numerically) shown to generate stripes and spots, but never gaps \cite{Fuentes2003,Hernandez-Garcia2004,martines2013vegetation,martinez-garcia2014}. We explain these early results analytically. Spots are ubiquitous in these competition models, and gaps never appear because the condition that drives the instability, $\hat{\phi}_p(k_c)<0$, forbids them. 
Stripes are also forbidden for the entire family of super-Gaussian competition kernels that can induce a Turing instability. They could, however, emerge for more complex interaction kernels, explaining previous numerical observations.
Higher-order competition eliminates both restrictions, providing a minimal demographic feedback that recovers the full diversity of spots, stripes and gaps. This form of interaction arises naturally in populations competing for a shared pool of resources \cite{Mendez2019,Jorge2023}, and it makes the competitive pressure felt by an individual exceed the sum of its pairwise interactions. A single non-additive negative feedback is therefore sufficient to explain morphological diversity.

Because we propose a competition density-dependence that can be measured at the individual level, our results provide a prediction that can be tested in the field. In dispersing populations with competition-driven patterns, pairwise interactions produce hexagonally arranged spots and nothing else within the canonical pattern family under typical competition kernels. Any other morphology in that family therefore strongly suggests higher-order competition or intricate competition kernels. Such a test requires the pattern geometry and qualitative evidence that competition is the pattern-forming feedback, without a direct estimate of $\gamma$. Dryland vegetation offers a natural setting to conduct such a test. Spots, stripes, and gaps have all been reported in these ecosystems \cite{Rietkerk2008, VonHardenberg2001}, remotely sensed imagery is abundant and easily accessible \cite{kastner_unravelling_2024}, and recently proposed experimental approaches could discriminate whether patterns arise from a scale-dependent activation-inhibition feedback or from competition alone \cite{martinez2023integrating}.

The phase diagram we constructed using nonlinear analysis (Fig.\,\ref{F3}) also shows non-standard relationships between the control parameter and pattern shape. Specifically, near $d_0/b \simeq 0.55$, increasing the intensity of higher-order competition produces the unusual sequence spots\,$\rightarrow$\,stripes\,$\rightarrow$\,spots\,$\rightarrow$\,uniform, even though $\kappa$ stays close to an increasing linear function of $\gamma$, which alone would typically give a single transition from spots to gaps. This re-entrance comes from the nonlinear modulation that higher-order competition imposes on the cubic saturation coefficients, which reshapes the stability domain of each morphology. Spots therefore appear on both sides of a stripe band, providing a counterexample to the standard uniform\,$\rightarrow$\,gaps\,$\rightarrow$\,stripes\,$\rightarrow$\,spots sequence common in ecological models \cite{pinto-ramos_aperiodic_2025,ruiz2020general}. This result indicates that spot patterns are not an unambiguous indicator of proximity to a tipping point \cite{rietkerk2004self,rietkerk2021evasion}.

In summary, we showed that competition alone can cause a single population to self-organize into spots, stripes, and gaps, provided the competition is non-additive. These results establish a new theoretical baseline for modeling spatial population dynamics, including the mechanistic understanding of how observed patterns emerge and the development of reliable early-warning signals of population collapse.

\textit{Data Availability Statement--}The code used to generate and process the data underlying the results reported in this work is publicly available at \url{https://github.com/DPintoRamos/HOI_Competing_Populations} and archived in Zenodo under the DOI \url{https://doi.org/10.5281/zenodo.22937226} \cite{pinto_ramos_2026_22937227}.
\begin{acknowledgments}
\textit{Acknowledgments--}This work was partially funded by the Center of Advanced Systems Understanding (CASUS), which is financed by Germany’s Federal Ministry of Research, Technology and Space (BMFTR) and by the Saxon Ministry for Science, Culture and Tourism (SMWK) with tax funds on the basis of the budget approved by the Saxon State Parliament. RMG also received support from the São Paulo Research Foundation (FAPESP) through ICTP-SAIFR 2021/14335-0.
\end{acknowledgments}

\bibliographystyle{apsrev4-2}
\bibliography{Vegetation_curated}

\onecolumngrid
\vspace{1cm}

\newpage
\section*{Supplementary Materials}

\setcounter{secnumdepth}{4}

\setcounter{section}{0}
\setcounter{equation}{0}
\setcounter{figure}{0}
\setcounter{table}{0}
\renewcommand{\thesection}{S\arabic{section}}
\renewcommand{\theequation}{S\arabic{equation}}
\renewcommand{\thefigure}{S\arabic{figure}}
\renewcommand{\thetable}{S\arabic{table}}

\makeatletter
\let\oldaddcontentsline\addcontentsline
\renewcommand{\addcontentsline}[3]{%
  \def\target{#1}\def\maintoc{toc}%
  \ifx\target\maintoc
    \oldaddcontentsline{toc_apendice}{#2}{#3}%
  \else
    \oldaddcontentsline{#1}{#2}{#3}%
  \fi
}
\makeatother

\makeatletter
\@starttoc{toc_apendice}
\makeatother

\newpage
\section{Implementaiton of the individual-based model simulations}

The continuous-time spatial birth-death process was simulated using a fixed-step, mortality-first, $\tau$-leaping approximation. The density-dependent mortality rates were updated every $m=10$ steps. Individuals were simulated in a square periodic domain of length $L$. The implementation proceeds as follows:

\begin{enumerate}

    \item The simulation is initialized with $N_0=4\times10^3$ individuals with positions sampled from a uniform distribution over the domain. Two arrays, $\mathbf{y}$ and $\mathbf{d}$, track, for each individual $i$, the position $(x,y)_i$ and the death-rate contributions $(d_0,d_p,d_{H.O.})_i$. The arrays have a preallocated length of $N_{\text{prealloc}}=10^6$, which was always greater than the total population in our simulations (steady-state populations between $1.0\times 10^5$ and $2.5 \times 10^5$). An integer $N_{\text{pop}}$ tracks the current population size, corresponding to the first $N_{\text{pop}}$ entries of the $\mathbf{y}$ and $\mathbf{d}$ arrays.

    \item To accelerate the calculation of the local densities determining the individual death rates, the spatial domain is partitioned into a regular grid of cells. The cells are created using the largest interaction distance, in our case $r_c$. Specifically, we define a partition of the square domain of length $L$ into
    \[
        n_{\text{cell}}=\lfloor L/r_c \rfloor
    \]
    intervals in each direction, giving square cells of length
    \[
        L_{\text{cell}}=L/n_{\text{cell}}.
    \]

    \item Assign each individual to a cell. Auxiliary arrays count the number of individuals in each cell and store the individual indices in a separate array, such that indices belonging to the same cell occupy contiguous positions.

    \item The array $\mathbf{d}$ is initialized by sweeping over every individual in parallel. For each individual, the number of neighbors within the radii $r_c$ and $r_{h.o.}$ is needed. For this, we exploit the cell auxiliary arrays, so we only need to search for individuals in the focal individual's cell and its 8 direct neighboring cells. The numbers of interacting individuals within the two radii are stored in the variables $C_p$ and $C_{H.O.}$, and the values
    \[
        d_p=\frac{C_p}{\pi r_c^2}
    \]
    and
    \[
        d_{H.O.}
        =
        \gamma
        \left(
            \frac{C_{H.O.}}{\pi r_{h.o.}^2}
        \right)^2
    \]
    are stored for each particle $i$.

    \item Time evolution loop:

    \begin{enumerate}

        \item Each individual dies according to a Poisson process, considering its total death rate fixed within the interval $\tau$. The death probability is evaluated in parallel:
        \[
            P_{\text{death}}
            =
            1-e^{-(d_0+d_p+d_{H.O.})\tau},
        \]
        and a random number $u$, uniformly sampled from the interval $[0,1]$, determines the death of the respective particle if $u<P_{\text{death}}$.

        \item Dead individuals are swapped with individuals at the last populated indices of the arrays $\mathbf{y}$ and $\mathbf{d}$, and $N_{\text{pop}}$ is reduced accordingly. This avoids the relocation or complete rewriting of the arrays $\mathbf{y}$ and $\mathbf{d}$.

        \item After the mortality process, surviving individuals are allowed to reproduce. According to the Poisson process, each particle produces a number of offspring $N_{\text{offspring}}$ drawn from a Poisson distribution with parameter $b\tau$. Each offspring is randomly placed according to $\phi_D$, considering the periodic boundary conditions, and initially inherits the death rates $\mathbf{d}_i$ of its parent. Newborn individuals enter the demographic process in the following $\tau$ step.

        \item Every $m=10$ iterations, the cell lists and the array $\mathbf{d}$ are recalculated for the whole population following the details of step 4.

    \end{enumerate}

\end{enumerate}
This implementation was written in Python, and partially compiled and accelerated with the Numba library.

\newpage
\section{Derivation of the density field equation}\label{sec:SM-meanfield}
We start by considering the birth-death rules: i) An individual at position $\mathbf{r}$ gives birth, with fixed birth-rate $b$, and the offspring is placed randomly in a position $\mathbf{r'}$ according to a probability density $\phi_D(\mathbf{r}-\mathbf{r'})$. ii) An individual at position $\mathbf{r}$ can die with a fixed baseline death-rate, $d_0$, and a density dependent rate, $d_c$, that depends on the surrounding individuals due to competition, with $d_c$ reading
\begin{eqnarray}
d_c(\mathbf{r}) = \eta\sum_{i}^{N} \phi_{p}(\mathbf{r}-\mathbf{r_i}) + \gamma \sum_{i}^N\sum_j^N \psi_{\mathrm{\tiny{H.O.}}}(\mathbf{r}, \mathbf{r_i}, \mathbf{r_j}).
\end{eqnarray}

We next discretize the space into a grid with $M\times M$ cells, each with an area $dx^2=L^2/M^2$. Each of these cells is labeled by $\mathbf{i}=(i, j)$ and contains a number of individuals $N_{\mathbf{i}}$, such that the microstate of the system is $\mathbf{\Omega}=\{N_{1,1}, N_{1,2}, ..., N_{M,M}\}$. We consider the system to be a one-step process; that is, total population can increase or decrease only by one at each infinitesimal time increment, or in other words, the probability of two or more events (birth or death) occurring simultaneously vanishes. With those considerations, we can compute all the possible states after an infinitesimal time interval $dt$, corresponding to
$$ \mathbf{\Omega}_{k,l}^{\pm}=\{N_{1,1}, N_{1,2}, ..., N_{k,l}\pm1,..., N_{M,M}\}.$$

Now, we compute the transition rates from a state $\mathbf{\Omega}$ to states $\mathbf{\Omega}_{k,l}^\pm$. Considering the case in which population increases by one, one computes that every individual has a fixed probability per unit time of $b$ for giving birth. The probability that the offspring is born at $(k,l)$ from a parent at $(i,j)$ is exactly $b N_{i,j} \phi_D( \mathbf{r}_{k,l}-\mathbf{r}_{i,j}) dx^2$; this considers that all potential parents at $(k,l)$ contribute additively to the probability, and that the kernel density is multiplied by the area of the grid square at $(k,l)$ (as any position within the square counts toward increasing the population $N_{k,l}$). At this point, it is approximated that all the individuals inside a grid square $(i,j)$ have the same position $\mathbf{r}_{i,j}$. It must follow that the probability that parents at position $(i,j)$ give birth to offspring anywhere is $\sum_{k,l} b N_{i,j} \phi_D( \mathbf{r}_{k,l}-\mathbf{r}_{i,j}) dx^2 \approx b N_{i,j} \int \phi_D(\mathbf{r})\mathrm{d}\mathbf{r} = b N_{i,j}$, which is the expected normalization. Now, we need to consider that the offspring at $(k,l)$ could come from any point in the grid, then, the rate at which the state changes to $\mathbf{\Omega}_{k,l}^+$ is 
$$ W(\mathbf{\Omega} \to \mathbf{\Omega}_{k,l}^+) = \sum_{i,j} b N_{i,j} \phi_D( \mathbf{r}_{k,l}-\mathbf{r}_{i,j}) dx^2.$$ 

We apply the same logic to deaths. Each individual has a death probability per unit time of $d=d_0 + d_c$. Then, an individual at cell $(k,l)$ has an increased death rate due to neighbors with a strength according to the kernels. Recalling our approximation that all individuals within a cell share the same position $\mathbf{r}_{i,j}$, the probability that an individual at $(k,l)$ dies corresponds to $d_0 + \sum_{i,j}N_{i,j}\phi_p(\mathbf{r}_{k,l}-\mathbf{r}_{i,j})+\gamma\sum_{i,j}\sum_{m,n} N_{i,j}N_{m,n}\psi_{\mathrm{H.O.}}(\mathbf{r}_{k,l},\mathbf{r}_{i,j},\mathbf{r}_{m,n})$. As each individual at $(k,l)$ is independent, the total transition rate to the state $\mathbf{\Omega}_{k,l}^-$ is
$$ W(\mathbf{\Omega} \to \mathbf{\Omega}_{k,l}^-)= N_{k,l}\left(d_0 + \eta\sum_{i,j}N_{i,j}\phi_p(\mathbf{r}_{k,l}-\mathbf{r}_{i,j})+\gamma \sum_{i,j}\sum_{m,n} N_{i,j}N_{m,n}\psi_{\mathrm{H.O.}}(\mathbf{r}_{k,l},\mathbf{r}_{i,j},\mathbf{r}_{m,n}) \right).$$

The Master equation dictating how the probability of the microstate $\mathbf{\Omega}$ evolves over time reads 
$$\frac{\partial P}{\partial t}= \sum_\mathbf{\Omega'}W(\mathbf{\Omega'}\to \mathbf{\Omega})P(\mathbf{\Omega'})- W(\mathbf{\Omega}\to \mathbf{\Omega'})P(\mathbf{\Omega}),$$
which will have the following explicit contributions
$$\frac{\partial P}{\partial t}= \sum_{k,l} W(\mathbf{\Omega}_{k,l}^-\to \mathbf{\Omega})P(\mathbf{\Omega}_{k,l}^-)+W(\mathbf{\Omega}_{k,l}^+\to \mathbf{\Omega})P(\mathbf{\Omega}_{k,l}^+)-W(\mathbf{\Omega} \to \mathbf{\Omega}_{k,l}^-)P(\mathbf{\Omega})- W(\mathbf{\Omega} \to \mathbf{\Omega}_{k,l}^+)P(\mathbf{\Omega}).$$
Lastly, we want to obtain
$$\frac{\partial \langle N_{i,j} \rangle}{\partial t}= \int N_{i,j}\frac{\partial P}{\partial t} \mathrm{d}\mathbf{\Omega},$$
which, considering that $N_{i,j}$ is discrete, becomes a sum over all the possible populations

$$\frac{\partial \langle N_{i,j} \rangle}{\partial t}= \left( \prod_{m,n}^{M,M}\sum_{N_{m,n}=0}^{\infty}\right) N_{i,j}\frac{\partial P}{\partial t}=\sum_{N_{1,1}=0}^{\infty}\sum_{N_{1,2}=0}^{\infty}...\sum_{N_{M,M}=0}^{\infty}N_{i,j}\frac{\partial P}{\partial t}.$$

We evaluate the right-hand side for the birth processes and the death processes follow analogously:
$$\frac{\partial P}{\partial t}_{\text births}= \sum_{k,l} W(\mathbf{\Omega}_{k,l}^-\to \mathbf{\Omega})P(\mathbf{\Omega}_{k,l}^-)- W(\mathbf{\Omega} \to \mathbf{\Omega}_{k,l}^+)P(\mathbf{\Omega}),$$
We will need to evaluate the following term, corresponding to the first term of the right-hand side:
$$\left( \prod_{m,n}^{M,M}\sum_{N_{m,n}=0}^{\infty}\right) N_{i,j} \sum_{k,l} \sum_{p,q} b (N_{p,q}-\delta_{k,p}\delta_{l,q})  \phi_D( \mathbf{r}_{k,l}-\mathbf{r}_{p,q}) dx^2 P(\mathbf{\Omega}_{k,l}^-),$$
interchanging the summations becomes
$$=b\sum_{k,l} \sum_{p,q}\left( \prod_{m,n}^{M,M}\sum_{N_{m,n}=0}^{\infty}\right) N_{i,j} (N_{p,q}- \delta_{k,p}\delta_{l,q})\phi_D( \mathbf{r}_{k,l}-\mathbf{r}_{p,q}) dx^2 P(\mathbf{\Omega}_{k,l}^-). $$
Let us perform first the sum over $N_{k,l}$
$$\sum_{N_{k,l}=0}^\infty N_{i,j} (N_{p,q}- \delta_{k,p}\delta_{l,q})\phi_D( \mathbf{r}_{k,l}-\mathbf{r}_{p,q}) dx^2 P(\mathbf{\Omega}_{k,l}^-), $$
and note that we can use the following property:
$$ \sum_{n=0}^{N-1} g(n) f(n+1) = \sum_{n=1}^{N}g(n-1)f(n),$$
considering that $g(n)$ is, in our equations, a power of $n$, and $f(n)$ is the probability of the microstate times a function of $n$. Note that due to the boundary condition at $n=0$, the first term of the summation is always vanishing; similarly, for the upper limit the probability of infinitely many particles must vanish. With those considerations, our first sum over $N_{k,l}$ can be written as
$$= \sum_{N_{k,l}=0}^\infty (N_{i,j}+\delta_{k,i}\delta_{l,j}) N_{p,q}\phi_D( \mathbf{r}_{k,l}-\mathbf{r}_{p,q}) dx^2 P(\mathbf{\Omega}).$$
Finally, for all the birth processes contributing to the Master equation, we have
$$\frac{\partial \langle N_{i,j} \rangle}{\partial t}_{\text births}=b\left( \prod_{m,n}^{M,M}\sum_{N_{m,n}=0}^{\infty}\right) \sum_{k,l} \sum_{p,q} \phi_D( \mathbf{r}_{k,l}-\mathbf{r}_{p,q}) dx^2 P(\mathbf{\Omega}) \left[ (N_{i,j}+\delta_{k,i}\delta_{l,j})N_{p,q} - N_{i,j}N_{p,q} \right], $$
$$= b\left( \prod_{m,n}^{M,M}\sum_{N_{m,n}=0}^{\infty}\right) \sum_{k,l} \sum_{p,q} \phi_D( \mathbf{r}_{k,l}-\mathbf{r}_{p,q}) dx^2 P(\mathbf{\Omega}) N_{p,q} \delta_{k,i}\delta_{l,j},$$
$$= b \sum_{p,q} \phi_D( \mathbf{r}_{i,j}-\mathbf{r}_{p,q})\langle N_{p,q}\rangle dx^2.$$
The competition term follows analogously, obtaining
$$\frac{\partial \langle N_{i,j} \rangle}{\partial t}_{\text deaths}=- d_0 \langle N_{i,j}\rangle  - \eta\sum_{p, q} \phi_p(\mathbf{r}_{i,j}-\mathbf{r}_{p,q})\langle N_{i,j} N_{p,q} \rangle- \gamma \sum_{p,q}\sum_{r,s} \psi_{\mathrm{H.O.}}(\mathbf{r}_{i,j},\mathbf{r}_{p,q},\mathbf{r}_{r,s})\langle N_{i,j}N_{p,q}N_{r,s}\rangle.$$
Note that self-interaction terms are neglected as they become irrelevant in the high-population limit. The death term involves the higher moments of the random variables $N_{i,j}$. We assume the limit of high densities where a mean-field approximation is valid; that is $N_{i,j} \approx \langle N_{i,j} \rangle + \delta N_{i,j}$, with $\delta N_{i,j} \ll N_{i,j}$. Therefore, $\langle N_{i,j} N_{p,q} ...\rangle \approx \langle N_{i,j} \rangle \langle N_{p, q}\rangle ... $ and we obtain a closed equation for $\mathrm{d}\langle N_{i,j}\rangle /\mathrm{d}t$. Defining $n(\mathbf{r})=N_{i,j}/dx^2$ and $\sum_{i,j} f(N_{i,j})\approx1/dx^2\int f(n(\mathbf{x})dx^2)\mathrm{d}\mathbf{r}$ for small enough $dx$, one obtains:
$$\frac{\partial n}{\partial t}= b \int \phi_D(\mathbf{r}-\mathbf{r'})n(\mathbf{r'})\mathrm{d}\mathbf{r'}- d_0 n-\eta n\int \phi_p(\mathbf{r}-\mathbf{r'})n(\mathbf{r'})\mathrm{d}\mathbf{r'} - \gamma n \iint \psi_{\mathrm{H.O.}}(\mathbf{r},\mathbf{r'}, \mathbf{r''})n(\mathbf{r'})n(\mathbf{r''})\mathrm{d}\mathbf{r'}\mathrm{d}\mathbf{r''}.$$

\newpage
\section{Implementation of the density-field equation simulation}\label{sec:SM-num}
We employ a standard integration scheme that discretizes space and computes integrals in the corresponding Fourier space with a given resolution. More precisely: 

\begin{enumerate}
    \item We define a rectangular spatial grid of lengths $L_x$ and $L_y$ using uniform spacing $dx$. Simulations in square domains are shown in Fig.\,\ref{F1} with $L_x=L_y=10$ and $dx=0.05$. Simulations corresponding to Fig. \ref{F2} were made in rectangular domains of $L_x=dx N_x=20$ and $L_y=dx N_y \approx 2L_x/\sqrt{3}$, such that hexagonal symmetry is less frustrated by the square domain partition discretization, and $dx=0.1$. The number of spatial points is $N_x=L_x/dx$ and $N_y=\lfloor 2L_x/(\sqrt{3} dx)\rfloor$, respectively.
    \item With the spatial resolution given, Fourier coordinates are constructed as: $f_x=[-\frac{1}{2dx}, -\frac{N_x-1}{2N_xdx}, ..., \frac{N_x-1}{2N_xdx}]$ and $f_y= [-\frac{1}{2dx}, -\frac{N_yx-1}{2N_ydx}, ..., \frac{N_y-1}{2N_ydx}]$. Wavevectors $\mathbf{k}$ follow by including a factor of $2\pi$. With those, one computes numerically the Fourier transform of the kernels, $\hat{\phi}_D$, $\hat{\phi}_p$, $\hat{\phi}_{h}$.
    \item We initialize a \texttt{scipy.integrate.ode} instance. We employ the Dormand-Prince 5 algorithm, an adaptive-time, 5th-order explicit time integrator. We pass the demographic and interaction parameters, as well as the kernel Fourier transforms, as arguments.
    \item Time-integration loop:
    \begin{enumerate}
        \item A Fourier transform of the field  $n(\mathbf{r}; t)$ yields $\hat{n}(\mathbf{k}; t)\equiv \mathcal{F}[n]$.
        \item The dispersal and interaction terms are computed using their convolution-like structure, such that
        $$\int \phi_D(\mathbf{r}-\mathbf{r'})n(\mathbf{r'})\mathrm{d}\mathbf{r'}= \mathcal{F}^{-1}\left[ \hat{\phi}_D(\mathbf{k}) \hat{n}(\mathbf{k}) \right],$$
        and analogously for the other terms.
        \item We compute the right-hand side in this way at each time step, packing the result in a single row-wise vector for the \texttt{scipy.integrate.ode} instance. At small sampling intervals $dt$, we extract snapshots of the density field.
    \end{enumerate}
    \item For Fig. \ref{F1}, each parameter set was initialized with a random initial condition and integrated until the dimensionless time $\tau_f\equiv bt_f=12800$.
    \item For Fig. \ref{F2}, we performed a continuation-like simulation protocol. Starting from $\gamma=0$, the simulation ran until $\tau_f=5000$ or up to the time such that every spatial point satisfied $|n(\mathbf{r}, t+dt)-n(\mathbf{r}, t)|<dt \times \text{Tol}$, and we used $\text{Tol}=10^{-5}$. After that, $\gamma$ is increased and the process starts again using as initial condition the last snapshot of the previous $\gamma$ value plus a small perturbation to shake the system out of metastable solutions.
\end{enumerate}

\newpage
\section{The anisotropy index and pattern amplitude}\label{sec:SM-anisotropy}
We use an anisotropy index to detect and systematically quantify the morphological changes in observed spatial patterns. We computed it with the following procedure:

\begin{enumerate}
    \item For each parameter, the density field must be recovered.
    \begin{description}
        \item[DFE] A discrete $n(\mathbf{r}, t)$ is obtained directly from the simulations.
        \item[IBM] The discrete density field must be reconstructed from the continuous positions of the individuals, $\mathbf{y}$. We perform the following:
        \begin{itemize}
            \item We choose a spatial resolution $dx$ and construct the discrete Fourier coordinates $f_x$ and $f_y$ (see \ref{sec:SM-num}).
            \item Considering the definition of the density field
            $$ n^{e}= \sum_{i}^{N_{\text{pop}}} \delta(\mathbf{r}- \mathbf{y}_i),$$
            its Fourier transform reads
            $$ \hat{n}^{e}=  \sum_{i}^{N_{\text{pop}}} e^{\mathbf{i} \mathbf{k}\cdot \mathbf{y}_i}.$$
            \item The raw discretized density field is obtained from computing $\hat{n}^{e}$ in the discrete Fourier space (using $f_x$ and $f_y$) and applying the inverse transform. That is, the discretized density field reads
            $$n_{\text{raw}}(\mathbf{r},t)= \mathcal{F}^{-1}[\hat{n}^{e}(\mathbf{k})].$$
            \item This field $n_{\text{raw}}$ yields a very noisy signal dominated by  Salt \& Pepper noise. We thus perform different smoothing operations depending on the property of the field we want to measure:
            \begin{description}
                \item[Anisotropy index calculation] The field is smoothed with a Gaussian filter of radius $\sigma_f=0.25\equiv \sigma$ physical spatial units.
                \item[Pattern amplitude calculation] The field was first smoothed with a Median filter of radius $\sigma_m=0.25\equiv \sigma$ spatial units (effective for Salt \& Pepper noise), followed by a Gaussian filter of radius $\sigma_f= \sigma/5$. Aggressive filtering is needed to avoid amplitudes, $\text{max}(n)-\text{min}(n)$, with values inflated by the demographic fluctuations.
            \end{description}
            \item Upon filtering, we obtain a smoother density field $n(\mathbf{r}, t)$.
        \end{itemize}
    \end{description}
    \item We compute the structure tensor of our pattern images, corresponding to the exterior product of the image gradient smoothed out in a sliding window of size $\sigma_{\mathcal{A}}$. We use the \texttt{scikit.image} Python package implementation, which uses a Gaussian kernel, $\phi_{s}(\mathbf{r}-\mathbf{r'}; \sigma_{\mathcal{A}})$, as the smoothing function. More precisely:
    $$S(\mathbf{r})= \begin{pmatrix}
        \phi_s \star \left(\frac{\partial n}{\partial x}\right)^2 & \phi_s \star \left(\frac{\partial n}{\partial x}\right)\left(\frac{\partial n}{\partial y}\right) \\
        \phi_s \star \left(\frac{\partial n}{\partial x}\right)\left(\frac{\partial n}{\partial y}\right) & \phi_s \star \left(\frac{\partial n}{\partial y}\right)^2
    \end{pmatrix},$$
    where $\star$ denotes the convolution operation.
    \item Eigenvalues of the structure tensor, $l_1$ and $l_2$, indicate whether there is a dominant direction in the rate-of-change of the image. Balanced eigenvalues, $l_1 \sim  l_2$, indicate no dominant direction, as the two orthogonal eigenvectors change at a similar rate. Imbalanced eigenvalues, by contrast, indicate a well-defined gradient direction. The local anisotropy index is defined as $\mathcal{A}_l(\mathbf{r})=|l_1-l_2|/(l_1+l_2)$.
    \item We report, for each parameter set, the median across realizations of the anisotropy index, corresponding to the median across the image of the local anisotropy index. That is, for each simulation realization $k$ $\mathcal{A}^k=\text{Median}[\mathcal{A}_l^k(\mathbf{r})]$, and we report $\mathcal{A}= \text{Median}[\mathcal{A}^k]$. For the DFE we use 32 independent initial conditions. For the IBM we use 10 independent realizations with 20 snapshots each.
    \item The smoothing distance, $\sigma_\mathcal{A}=2.5$ physical spatial units, is chosen to be at least bigger than the pattern characteristic wavelength. This choice ensures that the smoothing windows contain at least one spot or gap and that the gradient-weighted average vanishes, but for stripes it does not.
\end{enumerate}

\newpage
\section{Derivation of the pattern amplitude equations}\label{sec:SM-ampl}
Recall that we fixed $\eta=1$ and defined, for simplicity, $\psi_{\mathrm{H.O.}}(\mathbf{r},\mathbf{r'},\mathbf{r''})=\phi_{h.o.}(\mathbf{r}-\mathbf{r'})\phi_{h.o.}(\mathbf{r}-\mathbf{r''})$, with $\phi_{h.o.}$ a top-hat kernel of radius $r_{h.o.}$. Therefore, our equation reads
\begin{eqnarray}
    \frac{\partial n}{\partial t}= b \int \phi_D(\mathbf{r}-\mathbf{r'})n(\mathbf{r'})\mathrm{d}\mathbf{r'}- d_0 n-n\int \phi_p(\mathbf{r}-\mathbf{r'})n(\mathbf{r'})\mathrm{d}\mathbf{r'} - \gamma n \left[\int \phi_{h.o.}(\mathbf{r}-\mathbf{r'})n(\mathbf{r'})\mathrm{d}\mathbf{r'}\right]^2
    \label{Eq_den}
\end{eqnarray}

We start by computing the homogeneous state and its stability against heterogeneous perturbations. The homogeneous state follows from the equation
$$ 0 = (b - d_0)n - n^2  -\gamma n^3.$$
The unpopulated state, $n=0$, is a solution. The remaining solutions are
$$n= \frac{-1 \pm \sqrt{1+4\gamma(b-d_0)}}{2\gamma},$$
with the positive root being the only one that is physically relevant. We call the positive root $n_0$, and we let it be a function of $\gamma$, $n_0=n_0(\gamma)$. Writing Eq. \eqref{Eq_den} using $n_0(\gamma)$ as the reference state; that is, replacing $n(\mathbf{r},t)=n_0(\gamma)+u(\mathbf{r},t)$ leads to (note that the arguments are omitted where possible)
\begin{eqnarray}
    \frac{\partial  u}{\partial t}= && b\left(\int \phi_D(\mathbf{r-r'})u(\mathbf{r'})\mathrm{d}\mathbf{r'} - u\right)- n_0\int\phi_p(\mathbf{r-r'})u(\mathbf{r'})\mathrm{d}\mathbf{r'}-2\gamma n_0^2\int \phi_{h.o.}(\mathbf{r-r'})u(\mathbf{r'})\mathrm{d}\mathbf{r'} -\nonumber \\
   && u\int\phi_p(\mathbf{r-r'})u(\mathbf{r'})\mathrm{d}\mathbf{r'}-2\gamma n_0 u \int \phi_{h.o.}(\mathbf{r-r'})u(\mathbf{r'})\mathrm{d}\mathbf{r'}- \gamma n_0 \left(\int \phi_{h.o.}(\mathbf{r-r'})u(\mathbf{r'})\mathrm{d}\mathbf{r'}\right)^2-\nonumber \\
    &&\gamma u \left(\int \phi_{h.o.}(\mathbf{r-r'})u(\mathbf{r'})\mathrm{d}\mathbf{r'}\right)^2. \label{Eq_den_n0}
\end{eqnarray}
We note that Eq.\eqref{Eq_den_n0} can be written as
\begin{eqnarray}
    \frac{\partial  u}{\partial t}= \mathcal{L}u + F[u]^{[2]} + F[u]^{[3]}, \label{problem_0}
\end{eqnarray}
where $\mathcal{L}$ is a linear operator and $F[u]^{[n]}$ are nonlinearities of power $n$ in $u$. Neglecting the nonlinear terms, one can perform a linear stability analysis; for this, Fourier transform Eq.\eqref{problem_0}, obtaining
$$ \frac{\partial  \hat{u}}{\partial t}= b (\hat{\phi}_D-1)\hat{u} - n_0 \hat{\phi}_p\hat{u} - 2\gamma n_0^2 \hat{\phi}_{h.o.} \hat{u},$$
where
$$\hat{\phi}_D = e^{-\frac{s^2 \mathbf{k}^2}{2}},$$
$$\hat{\phi}_p= 2\frac{J_1(r_c |\mathbf{k}|)}{r_c |\mathbf{k}|},$$
and
$$\hat{\phi}_{h.o.}= 2\frac{J_1(r_{h.o.} |\mathbf{k}|)}{r_{h.o.} |\mathbf{k}|}.$$
Solving for $\hat{u}\propto e^{\lambda(\mathbf{k})t}$ one obtains (note that the dependence on $b$ is not written, as our control parameter is $\gamma$)
$$\lambda(\mathbf{k}, \gamma)\equiv\lambda(|\mathbf{k}|=k, \gamma)= b (\hat{\phi}_D-1) - n_0 \hat{\phi}_p - 2\gamma n_0^2 \hat{\phi}_{h.o.},$$
which can become positive for high enough $n_0$ at the critical wavenumber $|\mathbf{k}| = k_c$. This sets the threshold for a pattern-forming instability. Considering $\gamma$ as our control parameter, it is straightforward that reducing $\gamma$ up to a threshold $\gamma_c$ can trigger this instability. In what follows, we assume these constants, $\gamma_c$ and $k_c$, as given. At $\gamma_c$, it is satisfied that $\lambda(\mathbf{k}=k_c \hat{k}, \gamma=\gamma_c)=0$ for any direction $\hat{k}$; this means that any combination of spatial modes with the same wavenumber can in principle emerge. We focus on the first nontrivial resonant combination, $\mathbf{k}_1+\mathbf{k}_2+\mathbf{k}_3=0$, where $|\mathbf{k}_1|=|\mathbf{k}_2|=|\mathbf{k}_3|=k_c$. With these slow modes, we propose a polynomial expansion in their amplitudes, $A_i$ ($i\in[1,2,3]$), for a slow-manifold reduction. More precisely, we let
\begin{eqnarray}
    u(\mathbf{r}, t ) = A_1(t) e^{\mathbf{i} \mathbf{k}_1\cdot\mathbf{r}}+A_2(t) e^{\mathbf{i} \mathbf{k}_2\cdot\mathbf{r} }+ A_3(t) e^{\mathbf{i} \mathbf{k}_3\cdot\mathbf{r} } + \text{c.c.} + 
    +u^{[2]} + u^{[3]}+ ...,\label{ansatz_u}
\end{eqnarray}
where $\text{c.c.}$ denotes the complex conjugate of the previous terms. The temporal evolution of the amplitudes has a similar expansion
\begin{eqnarray}
    \frac{\mathrm{d} A_i}{\mathrm{d}t}=f_i^{[1]}+f_i^{[2]}+f_i^{[3]}+....\label{ansatz_A}
\end{eqnarray}
Recognizing the space of eigenfunctions of the linear problem as the Fourier basis, we impose that nonlinear corrections to the state variable are not in the critical subspace $\Psi=\text{span}\{\psi_1,\psi_{-1}, \psi_2, \psi_{-2}, \psi_3, \psi_{-3}\}$, with $\psi_{\pm i}= e^{\pm \mathbf{i} \mathbf{k}_i \cdot \mathbf{r}}$. More precisely, $P u^{[m]}=0$ for $m\geq2$, with $Pv=\sum_{i=1}^3 
[\psi_i\langle\psi_i,v\rangle + \psi_{-i}\langle\psi_{-i},v\rangle]$ under the canonical inner product $\langle v(\mathbf{r}), w(\mathbf{r})\rangle = \frac{1}{|V|}\int_{V} \bar{v}(\mathbf{r}) w(\mathbf{r}) d\mathbf{r}$, where $V$ denotes the domain. It is important to note the property that $\psi_1\psi_2=\psi_{-3}$ and $\psi_{-1}\psi_{-2}=\psi_{3}$ for any cyclic permutation of the indices, that $\bar{\psi}_i=\psi_{-i}$, and $A_{-i}=\bar{A}_i$ is the condition for $u$ to be real, where $\overline{(\cdot)}$ denotes the complex conjugate.
\newline

In what follows, we will systematically find the expressions for $u^{[m]}$ and $f_i^{[m]}$ by replacing Eqs. \eqref{ansatz_u} and \eqref{ansatz_A} into Eq. \eqref{problem_0}, equating terms of the same order $[m]$, and solving them hierarchically. We start by computing
$$ \frac{\mathrm{d} u}{\mathrm{d}t}= \sum_{i=1}^{3} \left(\frac{\partial u}{\partial A_i} \frac{\mathrm{d} A_i}{\mathrm{d}t}+ \frac{\partial u}{\partial A_{-i}}\frac{\mathrm{d} A_{-i}}{\mathrm{d} t}\right)= \sum_{i=1}^{3} \left[ \left( \psi_i + \sum_{m>1} \frac{\partial u^{[m]}}{\partial A_i}\right)\sum_{m\geq1} f_i^{[m]} +\left( \psi_{-i} + \sum_{m>1} \frac{\partial u^{[m]}}{\partial A_{-i}}\right) \sum_{m\geq1} f_{-i}^{[m]}\right],$$
corresponding to the left-hand-side of Eq. \eqref{problem_0}. An example of the right-hand side would be the following term
$$u \int \phi_p(\mathbf{r}-\mathbf{r'} )u(\mathbf{r'})\mathrm{d}\mathbf{r'}=\left[\sum_{i=1}^3 A_i \psi_i+ \text{c.c.}+\sum_{m>1}u^{[m]} \right]\int \phi_p(\mathbf{r}-\mathbf{r'} ) \left[\sum_{i=1}^3 A_i \psi_i(\mathbf{r'})+ \text{c.c.}+ \sum_{m>1}u^{[m]}(\mathbf{r'}) \right]\mathrm{d}\mathbf{r'},$$
$$= \left[\sum_{i=1}^3 A_i \psi_i+ \text{c.c.}+ \sum_{m>1}u^{[m]} \right]\left(\sum_{i=1}^3 A_i \psi_i \hat{\phi}_p(\mathbf{k}_i) + \text{c.c.}+ \sum_{m>1}\int \phi_p(\mathbf{r}-\mathbf{r'} ) u^{[m]}(\mathbf{r'})\mathrm{d}\mathbf{r'} \right).$$
In practice, the functions $u^{[m]}$ will be polynomials of the critical modes times the amplitudes, $A_{\pm i}\psi_{\pm i}$; therefore, the last term in the previous expression will become a summation over different amplitude polynomials weighted by the Fourier transforms of the kernel evaluated at different wavenumbers. The same applies to all the other nonlinear terms on the right-hand side of Eq. \eqref{problem_0}, which are cumbersome to write. Let us start with the leading-order term,
\newline

\underline{$O(A^1 )$}:
\newline

Eq. \eqref{problem_0} at this order reads
$$\sum_{i=1}^3\left( \psi_i f_i^{[1]}+ \psi_{-i} f_{-i}^{[1]}\right) = \mathcal{L} \left( \sum_{i=1}^3 \left(A_i\psi_i + A_{-i}\psi_{-i}\right)\right). $$
As $\mathcal{L} \psi_i=\lambda(k_c, \gamma)\psi_i$ and the functions $\psi_i$ are orthogonal, we find that $$f_i^{[1]}=\lambda(k_c, \gamma) A_i\equiv \mu(\gamma) A_i.$$ This is expected, as the lowest order corresponds just to the linear problem.
\newline

\underline{$O(A^2)$}:
\newline

Collecting the terms in Eq. \eqref{problem_0} up to this order leads to
$$\sum_{i=1}^3 \left(\psi_i f_i^{[2]}+\psi_{-i} f_{-i}^{[2]}+ \frac{\partial u^{[2]}}{\partial A_i}f_i^{[1]}+\frac{\partial u^{[2]}}{\partial A_{-i}}f_{-i}^{[1]}\right) = \mathcal{L} u^{[2]}+ F[u]^{[2]}.$$
We observe a linear inhomogeneous problem for $u^{[2]}$, which has the homological operator, $$\mathcal{H}_\gamma=\mathcal{L}-\lambda(k_c, \gamma)\sum_i \left(A_i\frac{\partial}{\partial A_i}+ A_{-i}\frac{\partial}{\partial A_{-i}}\right),$$ acting on it. To explicitly solve for the two unknowns, one first needs to write $F[u(A_1, A_2,...)]^{[2]}$. Only quadratic terms in $u$ can generate such contributions; by looking at Eq. \eqref{Eq_den_n0}, one has three potential contributors. We develop each of them separately:
\begin{eqnarray*}
    \left(u\int\phi_p(\mathbf{r-r'})u(\mathbf{r'})\mathrm{d}\mathbf{r'}\right)^{[2]}=&&\left(\sum_{i=1}^3 A_i\psi_i + \overline{A_i \psi_i}\right)\left(\sum_{i=1}^3 A_i \psi_i \hat{\phi}_p(\mathbf{k_i}) + \overline{A_i \psi_i} \hat{\phi}_p(-\mathbf{k_i})\right),\\
    =&&\hat{\phi}_p(k_c) \left(\sum_{i=1}^3 A_i\psi_i + \overline{A_i \psi_i}\right)^2,\\
    =&& \hat{\phi}_p(k_c) \left(\sum_{i, j} A_iA_j\psi_i\psi_j+\overline{A_iA_j\psi_i\psi_j}+ 2A_i\psi_i\overline{A_j\psi_j} \right),\\
    =&& \hat{\phi}_p(k_c) \Bigg(\left[\sum_{i=1}^3 A_i^2\psi_i^2+\overline{A_i^2\psi_i^2}\right]  + 2(A_1A_2\psi_{-3}+ A_2A_3\psi_{-1}+A_3A_1\psi_{-2}+\text{c.c.})+\\ 
    && 2(A_1\bar{A}_2\psi_1\psi_{-2}+ A_2\bar{A}_3\psi_2\psi_{-3}+ A_3\bar{A}_1\psi_3\psi_{-1}+\text{c.c.})+ 2(|A_1|^2+|A_2|^2+|A_3|^2) \Bigg),
\end{eqnarray*}
where we have used the fact that $\hat{\phi}(\pm\mathbf{k}_i)=\hat{\phi}(|\mathbf{k}_i|)=\hat{\phi}(k_c)$ and $\psi_1\psi_2=\psi_{-3}$ for any cyclic permutation. Note that we have terms like $\psi_1\psi_{-2}=e^{\mathbf{i}(\mathbf{k}_1 - \mathbf{k}_2)\cdot \mathbf{r}}$, which have wavenumbers different than $k_c$. This will be important in the next order. For the current order, the terms like $\overline{A_1 A_2}\psi_1$ are relevant because they are in the subspace $\Psi$. The rest of the quadratic contributions have very similar expressions:
\begin{eqnarray*}
    \left(2\gamma n_0 u\int\phi_{h.o.}(\mathbf{r-r'})u(\mathbf{r'})\mathrm{d}\mathbf{r'}\right)^{[2]}= &&2\gamma n_0\hat{\phi}_{h.o.}(k_c) \Bigg(\left[\sum_{i=1}^3 A_i^2\psi_i^2+\overline{A_i^2\psi_i^2}\right]  + 2(A_1A_2\psi_{-3}+ A_2A_3\psi_{-1}+A_3A_1\psi_{-2}+\text{c.c.})+\\ 
    && 2(A_1\bar{A}_2\psi_1\psi_{-2}+ A_2\bar{A}_3\psi_2\psi_{-3}+ A_3\bar{A}_1\psi_3\psi_{-1}+\text{c.c.})+2(|A_1|^2+|A_2|^2+|A_3|^2) \Bigg),
\end{eqnarray*}
\begin{eqnarray*}
    \left(\gamma n_0 \left[\int\phi_{h.o.}(\mathbf{r-r'})u(\mathbf{r'})\mathrm{d}\mathbf{r'}\right]^2\right)^{[2]}= &&\gamma n_0[\hat{\phi}_{h.o.}(k_c)]^2 \Bigg(\left[\sum_{i=1}^3 A_i^2\psi_i^2+\overline{A_i^2\psi_i^2}\right]  + 2(A_1A_2\psi_{-3}+ A_2A_3\psi_{-1}+A_3A_1\psi_{-2}+\text{c.c.})+\\ 
    && 2(A_1\bar{A}_2\psi_1\psi_{-2}+ A_2\bar{A}_3\psi_2\psi_{-3}+ A_3\bar{A}_1\psi_3\psi_{-1}+\text{c.c.})+2(|A_1|^2+|A_2|^2+|A_3|^2) \Bigg).
\end{eqnarray*}
Finally, the inhomogeneous linear problem reads
\begin{eqnarray*}
    \sum_{i=1}^3 \left(\psi_i f_i^{[2]}+\psi_{-i} f_{-i}^{[2]}\right) = && \mathcal{H}_\gamma u^{[2]} -\left(\hat{\phi}_p(k_c)+2\gamma n_0\hat{\phi}_{h.o.}(k_c) + \gamma n_0 [\hat{\phi}_{h.o.}(k_c)]^2\right)\Bigg(\left[\sum_{i=1}^3 A_i^2\psi_i^2+\overline{A_i^2\psi_i^2}\right]  + \\
    &&2(A_1A_2\psi_{-3}+ A_2A_3\psi_{-1}+A_3A_1\psi_{-2}+\text{c.c.})+\\ 
    && 2(A_1\bar{A}_2\psi_1\psi_{-2}+ A_2\bar{A}_3\psi_2\psi_{-3}+ A_3\bar{A}_1\psi_3\psi_{-1}+\text{c.c.})+ \\
    && 2(|A_1|^2+|A_2|^2+|A_3|^2)\Bigg).
\end{eqnarray*}
By projecting on the subspace $\Psi$ noting that $P\mathcal{H_\gamma}u^{[m]}=\mathcal{H}_\gamma P u^{[m]}=0$, one arrives at
$$f_1^{[2]}= -2\left(\hat{\phi}_p(k_c)+2\gamma n_0\hat{\phi}_{h.o.}(k_c) + \gamma n_0 [\hat{\phi}_{h.o.}(k_c)]^2\right)\overline{A_2A_3}\equiv -\kappa(\gamma)\overline{A}_2\overline{A}_3,$$
its cyclic permutations, and its complex conjugates. The equation determining $u^{[2]}$, in the complement of $\Psi$, now reads
\begin{eqnarray*}
    \mathcal{H}_\gamma u^{[2]}= && \left(\hat{\phi}_p(k_c)+2\gamma n_0\hat{\phi}_{h.o.}(k_c) + \gamma n_0 [\hat{\phi}_{h.o.}(k_c)]^2\right)\Bigg(\sum_{i=1}^3 \left[A_i^2\psi_i^2+\overline{A_i^2\psi_i^2}\right]  + \\
    && 2(A_1\bar{A}_2\psi_1\psi_{-2}+ A_2\bar{A}_3\psi_2\psi_{-3}+ A_3\bar{A}_1\psi_3\psi_{-1}+\text{c.c.})+ \\
    && 2(|A_1|^2+|A_2|^2+|A_3|^2)\Bigg).
\end{eqnarray*}
Recalling that $$\mathcal{L} (\psi_m\psi_n...\psi_N)=\lambda(\mathbf{k}_m+\mathbf{k}_m+...+\mathbf{k}_N, \gamma)(\psi_m\psi_n...\psi_N),$$ and noting that $$\left[\sum_i \left( A_i\frac{\partial}{\partial A_i}+ A_{-i}\frac{\partial}{\partial A_{-i}}\right)\right] \prod_kA_k^{m_k}A_{-k}^{m_{-k}}= \left[\sum_i (m_i+m_{-i})\right]\prod_kA_k^{m_k}A_{-k}^{m_{-k}} ,$$ the solution can be found by inspection. We propose
\begin{eqnarray*}
    u^{[2]}=&&a_0(|A_1|^2+|A_2|^2+|A_3|^2) +\big(a_{1,1}\psi_1^2A_1^2+a_{2,2}\psi_2^2A_2^2+a_{3,3}\psi_3^2A_3^2 + a_{1,-2}\psi_1\psi_{-2}A_1\overline{A}_2+\\
    &&a_{2,-3}\psi_2\psi_{-3}A_2\overline{A}_3+a_{3, -1}\psi_3\psi_{-1}A_3\overline{A}_1+ \text{c.c.}\big),
\end{eqnarray*}
and upon inserting in the previous equation and grouping coefficients with the corresponding orthogonal functions, one obtains: 
$$a_0 = \frac{2\left(\hat{\phi}_p(k_c)+2\gamma n_0\hat{\phi}_{h.o.}(k_c) + \gamma n_0 [\hat{\phi}_{h.o.}(k_c)]^2\right) }{\lambda(k=0, \gamma)-2\lambda(k_c,\gamma)},$$
$$a_{i,i}= \frac{\left(\hat{\phi}_p(k_c)+2\gamma n_0\hat{\phi}_{h.o.}(k_c) + \gamma n_0 [\hat{\phi}_{h.o.}(k_c)]^2\right)}{\lambda(2k_c, \gamma)-2\lambda(k_c,\gamma)}, $$
and
$$a_{i, -j}=\frac{2\left(\hat{\phi}_p(k_c)+2\gamma n_0\hat{\phi}_{h.o.}(k_c) + \gamma n_0 [\hat{\phi}_{h.o.}(k_c)]^2\right) }{\lambda(\sqrt{3}k_c, \gamma)-2\lambda(k_c,\gamma)},$$
where we have used the fact that $|\mathbf{k}_1-\mathbf{k}_2|=|\mathbf{k_2}-\mathbf{k}_3|=|\mathbf{k}_3-\mathbf{k}_1|=\sqrt{2k_c^2-2\cos(2\pi/3)k_c^2}=\sqrt{3}k_c$.
\newline

\underline{$O(A^3)$}:
\newline

At this order, both quadratic (through terms $u^{[1]} u^{[2]}$) and cubic (through $(u^{[1]})^3$) terms will contribute to the right-hand-side term $F[u(A_1,A_2,..)]^{[3]}$ in Eq. \eqref{problem_0}. Similarly, on the left-hand-side terms such as $\partial u^{[2]}/\partial A_1 f_1^{[2]}$ appear at this order; nevertheless, those will be outside the subspace $\Psi$ and irrelevant for obtaining $f^{[3]}$. We note that the expressions for $f_i$ have been obtained by cyclic permutation of the indices, and that the linear inhomogeneous problem always has the same structure. Therefore, we will compute only for $f_1^{[3]}$ by tracking the terms proportional to $\psi_1$ in $F[u(A_1,A_2,...)]^{[3]}$ to avoid cumbersome expressions. One has
\begin{eqnarray*}
    f_1^{[3]}\psi_1 =&& \mathcal{H}_\gamma u^{[3]} - a_0(|A_1|^2+|A_2|^2+|A_3|^2)\left(1+ \hat{\phi}_p(k_c)+ 2\gamma n_0(1+\hat{\phi}_{h.o.}(k_c))  + 2\gamma n_0\hat{\phi}_{h.o.}(k_c)  \right)A_1\psi_1- \\
    &&a_{1,1}|A_1|^2A_1\left(\hat{\phi}_p(k_c)+\hat{\phi}_p(2k_c)+ 2\gamma n_0(\hat{\phi}_{h.o.}(k_c)+\hat{\phi}_{h.o.}(2k_c))  + 2\gamma n_0\hat{\phi}_{h.o.}(k_c)\hat{\phi}_{h.o.}(2k_c) \right)\psi_1-\\
    &&a_{1,-2}(|A_2|^2+|A_3|^2)A_1\left(\hat{\phi}_p(k_c)+\hat{\phi}_p(\sqrt{3}k_c)+ 2\gamma n_0(\hat{\phi}_{h.o.}(k_c)+\hat{\phi}_{h.o.}(\sqrt{3}k_c))  + 2\gamma n_0\hat{\phi}_{h.o.}(k_c)\hat{\phi}_{h.o.}(\sqrt{3}k_c) \right)\psi_1-\\
    &&\gamma [\hat{\phi}_{h.o.}(k_c)]^2 \left( 3|A_1|^2 + 6(|A_2|^2+|A_3|^2)  \right)A_1\psi_1 + \text{all other terms.}
\end{eqnarray*}
As before, $f_1^{[3]}$ will be such that it cancels all the terms proportional to $\psi_1$ on the right-hand side. We do not compute the correction $u^{[3]}$ as this is the last contribution we will consider. Finally, the equation for the amplitude reads
\begin{eqnarray}
    \frac{\mathrm{d}A_1}{\mathrm{d}t}= \mu(\gamma)A_1 - \kappa (\gamma)\overline{A}_2\overline{A}_3-c(\gamma)|A_1|^2A_1-\nu(\gamma)(|A_2|^2+|A_3|^2)A_1+O(A^4),
\end{eqnarray}
and the rest are obtained by cyclic permutation of the indices and conjugation. The coefficients $c$ and $\nu$ are explicitly
\begin{eqnarray*}
    c(\gamma)=&& 3\gamma [\hat{\phi}_{h.o.}(k_c)]^2 +a_0\left(1+ \hat{\phi}_p(k_c)+ 2\gamma n_0(1+\hat{\phi}_{h.o.}(k_c))  + 2\gamma n_0\hat{\phi}_{h.o.}(k_c)  \right)+\\
    &&a_{1,1}\left(\hat{\phi}_p(k_c)+\hat{\phi}_p(2k_c)+ 2\gamma n_0(\hat{\phi}_{h.o.}(k_c)+\hat{\phi}_{h.o.}(2k_c))  + 2\gamma n_0\hat{\phi}_{h.o.}(k_c)\hat{\phi}_{h.o.}(2k_c) \right),
\end{eqnarray*}
\begin{eqnarray*}
    \nu(\gamma)=&& 6\gamma [\hat{\phi}_{h.o.}(k_c)]^2 +a_0\left(1+ \hat{\phi}_p(k_c)+ 2\gamma n_0(1+\hat{\phi}_{h.o.}(k_c))  + 2\gamma n_0\hat{\phi}_{h.o.}(k_c)  \right)+\\
    &&a_{1,-2}\left(\hat{\phi}_p(k_c)+\hat{\phi}_p(\sqrt{3}k_c)+ 2\gamma n_0(\hat{\phi}_{h.o.}(k_c)+\hat{\phi}_{h.o.}(\sqrt{3}k_c))  + 2\gamma n_0\hat{\phi}_{h.o.}(k_c)\hat{\phi}_{h.o.}(\sqrt{3}k_c) \right).
\end{eqnarray*}

\newpage
\section{Pattern amplitudes and stability}

In what follows, we assume that the saturating coefficients $c$ and $\nu$ remain positive for the relevant domain of $\gamma$; this was verified numerically for all the parameters used throughout the work. 
\subsection{Pattern amplitude equilibria}
Solutions representing different patterns can be found from the amplitude equations. It will be useful to consider their complex form: 
\begin{eqnarray}
        \frac{\mathrm{d}A_1}{\mathrm{d}t}= \mu(\gamma)A_1 - \kappa (\gamma)\overline{A}_2\overline{A}_3-c(\gamma)|A_1|^2A_1-\nu(\gamma)(|A_2|^2+|A_3|^2)A_1, \nonumber \\
        \frac{\mathrm{d}A_2}{\mathrm{d}t}= \mu(\gamma)A_2 - \kappa (\gamma)\overline{A}_3\overline{A}_1-c(\gamma)|A_2|^2A_2-\nu(\gamma)(|A_3|^2+|A_1|^2)A_2, \nonumber \\
        \frac{\mathrm{d}A_3}{\mathrm{d}t}= \mu(\gamma)A_3 - \kappa (\gamma)\overline{A}_1\overline{A}_2-c(\gamma)|A_3|^2A_3-\nu(\gamma)(|A_1|^2+|A_2|^2)A_3, \nonumber \\
        \text{and their complex conjugates,}
        \label{complex_amps}
\end{eqnarray}
and their amplitude and phase representation by letting $A_i=\rho_i e^{\mathbf{i} \theta_i}$: 
\begin{eqnarray}
    \frac{\mathrm{d} \rho_1}{\mathrm{d}t}= \mu(\gamma)\rho_1 - \kappa(\gamma) \rho_2 \rho_3 \cos \Phi - c(\gamma) \rho_1^3 -\nu(\gamma)(\rho_2^2+\rho_3^2)\rho_1, \nonumber \\
    \frac{\mathrm{d} \rho_2}{\mathrm{d}t}= \mu(\gamma)\rho_2 - \kappa(\gamma) \rho_3 \rho_1 \cos \Phi - c(\gamma) \rho_2^3 -\nu(\gamma)(\rho_3^2+\rho_1^2)\rho_2, \nonumber \\
    \frac{\mathrm{d} \rho_3}{\mathrm{d}t}= \mu(\gamma)\rho_3 - \kappa(\gamma) \rho_1 \rho_2 \cos \Phi - c(\gamma) \rho_3^3 -\nu(\gamma)(\rho_1^2+\rho_2^2)\rho_3, \nonumber \\
    \frac{\mathrm{d} \Phi}{\mathrm{d}t}= \kappa(\gamma)\left( \frac{\rho_2 \rho_3}{\rho_1}+\frac{\rho_3 \rho_1}{\rho_2}+\frac{\rho_1 \rho_2}{\rho_3} \right)\sin \Phi, \nonumber \\
    \Phi= \theta_1+\theta_2+\theta_3, \nonumber \\
    \rho_i \neq 0.
    \label{amplitude-phase}
\end{eqnarray}
The last condition in Eq. \eqref{amplitude-phase} is required to reduce the phase equations to the single global phase $\Phi$.

\begin{description}
    \item[Stripe pattern] is found when only one mode dominates. Without loss of generality, consider $A_2=A_3=0$. Imposing the equilibrium condition in Eqs. \eqref{complex_amps}, the nontrivial equilibrium solution for a single mode is
    $$|A_1|= \sqrt{\frac{\mu(\gamma)}{c(\gamma)}}\equiv\rho_{\text{stripe}},$$
    and exists provided that $\mu(\gamma)>0$.
    \item[Hexagonal patterns] The three modes are equally relevant for the solution, so we let $\rho_1=\rho_2=\rho_3\equiv\rho$ in Eqs. \eqref{amplitude-phase}. Imposing the equilibrium condition, the nontrivial solution for $\rho$ is determined from
    $$0 = \mu(\gamma) - \kappa(\gamma) \cos(\Phi )\rho - [c(\gamma)+2\nu(\gamma)]\rho^2,$$
    obtaining two solutions
    $$\rho_{\pm}=\frac{-\kappa(\gamma)\cos\Phi \pm \sqrt{\kappa(\gamma)^2\cos^2\Phi +4\mu(\gamma)[c(\gamma)+2\nu(\gamma)]} }{2[c(\gamma)+2\nu(\gamma)]}.$$
    The global phase always has two fixed points, $\Phi=0$ and $\Phi=\pi$ (modulo $2\pi$). Their stability depends solely on the sign of $\kappa(\gamma)$. For $\kappa(\gamma)<0$, $\Phi=0$ is the stable equilibrium. For $\kappa(\gamma)>0$, $\Phi=\pi$ is the stable equilibrium. Hence $\cos^2\Phi=1$ at equilibrium. However, the stable equilibrium of $\Phi$ determines the sign of the first term in $\rho_{\pm}$.
    \begin{description}
        \item[Spots] Reconstructing the field at first order, $u = \rho_1e^{\mathbf{i}(\theta_1+\mathbf{k_1}\cdot \mathbf{x})}+\rho_2e^{\mathbf{i}(\theta_2+\mathbf{k_2}\cdot \mathbf{x})}+\rho_3e^{\mathbf{i}(\theta_3+\mathbf{k_3}\cdot \mathbf{x})}+\text{c.c.},$ one can see that for $\theta_i=0$, $\mathbf{x}=0$ reaches the maximum and the field decays outward from this central spot; this corresponds to a spot pattern. The general condition is that $\Phi=0$ (modulo $2\pi$). Therefore, spots \textit{may} form whenever $\kappa(\gamma)<0$. A condition for their general stability will be derived later. The amplitude for the spot patterns reads
        $$\rho_{\text{spot}}^{\pm}=\frac{-\kappa(\gamma)\pm \sqrt{\kappa(\gamma)^2 +4\mu(\gamma)[c(\gamma)+2\nu(\gamma)]} }{2[c(\gamma)+2\nu(\gamma)]}.$$
        These branches emerge from a saddle-node bifurcation at $4\mu(\gamma)[c(\gamma)+2\nu(\gamma)]=-\kappa(\gamma)^2$. However, this bifurcation is physically relevant only if $\mu(\gamma)<0$ at the same time as $\kappa(\gamma)<0$, otherwise, the negative branch remains below $0$.
        \item[Gaps] Similarly, if $\theta_i=\pi$, the reverse situation occurs: the field $u$ is a minimum at $\mathbf{x}=0$ and increases outward from this central gap, corresponding to a gap pattern. The general condition is $\Phi=\pi$ (modulo $2\pi$). Then, gap patterns \textit{may} occur when $\kappa(\gamma)>0$. The amplitude for the gaps reads
        $$\rho_{\text{gap}}^{\pm}=\frac{\kappa(\gamma)\pm \sqrt{\kappa(\gamma)^2 +4\mu(\gamma)[c(\gamma)+2\nu(\gamma)]} }{2[c(\gamma)+2\nu(\gamma)]}.$$
        Again, they emerge from a saddle-node bifurcation, which is only physically relevant if $\mu(\gamma)<0$ at the same time as $\kappa(\gamma)>0$. 
    \end{description}
\end{description}

\subsection{Stability thresholds}
The stability region of each pattern type, and thus the bistability regions shown in Fig. \ref{F2}, are derived from the eigenvalues of the linearized dynamics around the three pattern-type solutions of the amplitude equation. This information, together with the Turing instability threshold, $\gamma_c$, allows us to construct the full phase diagram of Fig. \ref{F3}. 
\begin{figure}[t!]
	\centering
	\includegraphics[width=\textwidth]{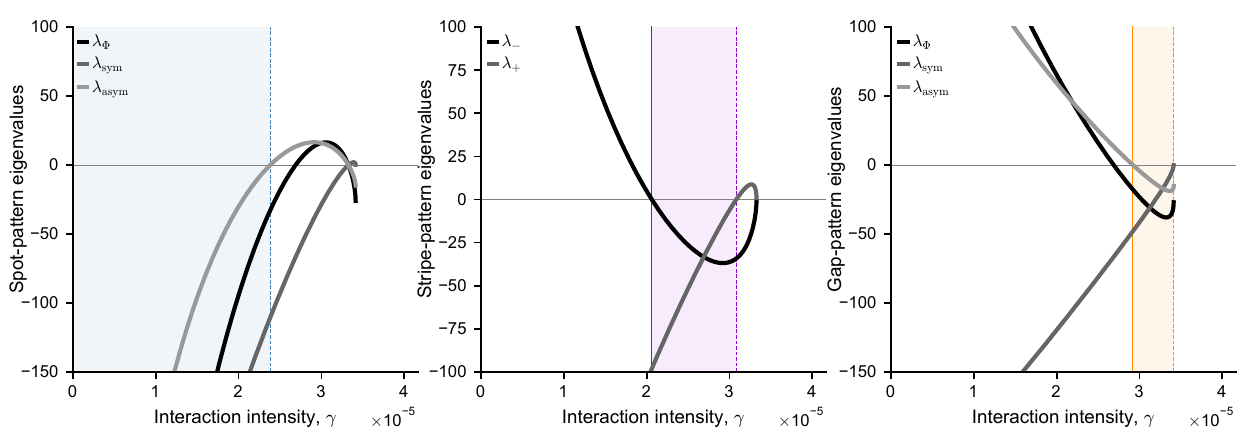}
	\caption{Eigenvalues of the three types of patterns. From left to right: spot patterns ($\Phi=0$), stripe patterns, and gap patterns ($\Phi=\pi$). The colored area indicates their region of stability derived from the eigenvalues of the linearized dynamics around them.}
	\label{FS1}
\end{figure}
\begin{description}
    \item[Stripes] We let $A_1= \rho_{\text{stripe}} e^{\mathbf{i} \theta_0} + \delta_1$, $A_2=\delta_2$, and $A_3= \delta_3$, with $|\delta_i| \ll 1$. Inserting these in Eqs. \eqref{complex_amps} and linearizing, one obtains three separable subsystems (function arguments are omitted):
    $$\frac{\mathrm{d}}{\mathrm{d}t}\begin{pmatrix}
        \delta_1 \\
        \bar{\delta}_1
    \end{pmatrix}= \begin{pmatrix}
        -\mu & -\mu e^{2\mathbf{i}\theta_0} \\
        -\mu e^{-2\mathbf{i}\theta_0}  & -\mu
    \end{pmatrix} \begin{pmatrix}
        \delta_1 \\
        \bar{\delta}_1
    \end{pmatrix},$$

       $$\frac{\mathrm{d}}{\mathrm{d}t}\begin{pmatrix}
        \delta_2 \\
        \bar{\delta}_3
    \end{pmatrix}= \begin{pmatrix}
        \mu(1-\frac{\nu}{c}) & -\kappa\sqrt{\frac{\mu}{c}} e^{-\mathbf{i}\theta_0} \\
        -\kappa\sqrt{\frac{\mu}{c}} e^{\mathbf{i}\theta_0}  & \mu(1-\frac{\nu}{c})
    \end{pmatrix} \begin{pmatrix}
        \delta_2 \\
        \bar{\delta}_3
    \end{pmatrix},$$

 $$\frac{\mathrm{d}}{\mathrm{d}t}\begin{pmatrix}
        \delta_3 \\
        \bar{\delta}_2
    \end{pmatrix}= \begin{pmatrix}
        \mu(1-\frac{\nu}{c}) & -\kappa\sqrt{\frac{\mu}{c}} e^{-\mathbf{i}\theta_0} \\
        -\kappa\sqrt{\frac{\mu}{c}} e^{\mathbf{i}\theta_0}  & \mu(1-\frac{\nu}{c})
    \end{pmatrix} \begin{pmatrix}
        \delta_3 \\
        \bar{\delta}_2
    \end{pmatrix},$$

From the first subsystem, one finds the eigenvalues
$$\lambda = 0,$$
$$\lambda=-2\mu.$$

The two identical remaining subsystems lead to the following eigenvalues with multiplicity 2:

$$\lambda_\pm= \mu(1-\frac{\nu}{c}) \pm \kappa\sqrt{\frac{\mu}{c}}.$$

The vanishing eigenvalue corresponds to the translation mode $\theta_0 \rightarrow \theta_0 + \delta\theta$. From the rest, one can deduce that the stripe pattern solution is stable if all the eigenvalues are negative. This leads to the conditions $\mu>0$, $\nu>c$, and $|\kappa \sqrt{\mu/c}|< |\mu(1-\nu/c)|$. It is easier to see the domain where this last condition is satisfied by plotting the nontrivial eigenvalues as a function of $\gamma$, as shown in Fig. \ref{FS1}.

\item[Hexagonal patterns] We use $\rho_+$ as the reference state, which turns out to be the stable branch (and for most of the $\gamma$ domain, the only physically sensible). Then, we let $\rho_i=\rho_+ + \omega_i$ and $\Phi= \Phi_0 + \theta$, with $\omega_i\ll1$ and $\theta\ll 1$, and insert them into Eqs. \eqref{amplitude-phase}. After linearizing, the equations for $\omega_i$ are independent of $\theta$, and one eigenvalue is trivially obtained: $$\lambda_\Phi= 3\rho_+\cos(\Phi_0) \kappa.$$ Two vanishing eigenvalues correspond to the translation of the hexagonal pattern. Lastly, the remaining eigenvalues are obtained from the amplitude subsystem. Defining
$$ a_\rho= \mu - 3c\rho_+^2-2\nu\rho_+^2,$$
$$b_\rho=-2\nu\rho_+^2- \kappa\rho_+\cos\Phi_0,$$
the linearized subsystem reads
 $$\frac{\mathrm{d}}{\mathrm{d}t}\begin{pmatrix}
        \omega_1 \\
       \omega_2 \\
       \omega_3
    \end{pmatrix}= \begin{pmatrix}
        a_\rho & b_\rho & b_\rho \\
        b_\rho  & a_\rho & b_\rho \\
        b_\rho & b_\rho & a_\rho
    \end{pmatrix} \begin{pmatrix}
        \omega_1 \\
       \omega_2 \\
       \omega_3
    \end{pmatrix}.$$
The characteristic polynomial is
$$(a_\rho-\lambda)^3 - 3(a_\rho-\lambda)b_\rho^2+ 2b_\rho^3=0,$$
which can be solved by letting
$$ (a_\rho-\lambda)= w+\frac{b_\rho^2}{w},$$
leading to 
$$(w^3)^2 + 2b_\rho^3w^3+ b_\rho^6=0.$$
Solving for $w$ and returning to the original variable leads to the eigenvalues
$$\lambda_\text{sym}= a_\rho +2b_\rho = -2c\rho_+^2- \kappa\rho_+\cos\Phi_0 - 4\nu \rho_+^2,$$
with multiplicity one, and
$$\lambda_\text{asym}=a_\rho-b_\rho=-2c\rho_+^2+2\kappa\rho_+ \cos\Phi_0+2\nu\rho_+^2,$$
with multiplicity two. In the previous equations, we have used the equation determining $\rho_+$ to reduce terms.

Now, the stability domain of spots and gaps depends on these eigenvalues. As we mentioned, the sign of $\kappa$ determines the stability of the $\Phi$ equilibria, while the amplitude stability is given by $\lambda_\text{sym}$ and $\lambda_\text{asym}$, as seen in Fig. \ref{FS1}.
\end{description}
\newpage

\section{Hexagonal spot restriction in the pairwise case}
In this section, we show that the limit with only pairwise competition restricts pattern selection to hexagonal spots within the canonical pattern family. This result holds for all demographic parameters and interaction ranges, and it requires Gaussian dispersal and a super-Gaussian competition kernel,     from which the top-hat is recovered as a limit:
$$\phi_p(\mathbf{x})= \frac{p}{2\pi r_c^2 \Gamma(2/p)} e^{-\left(\frac{|\mathbf{x}|}{r_c}\right)^p},$$
where $\Gamma(z)$ is the usual Gamma function. It is known that for $p>2$, the traditional nonlocal FKPP can suffer a Turing instability leading to patterns. A top-hat kernel corresponds to the case $p\to\infty$.

It is useful to write our amplitude equation parameters in the pairwise limit, $\gamma=0$. We define
\begin{eqnarray*}
    r&= &n_0(\gamma=0)= b-d_0, \\
    \beta&=&\frac{b}{r}=\frac{b}{b-d_0}\geq1, \\
    D_q&= &\hat{\phi}_D(q k_c),\\
    P_q&=&\hat{\phi}_p(q k_c),\\
    \mu_0&=&\mu(\gamma=0)= r[ \beta(D_1-1)-P_1] \\
    \kappa_0&=&\kappa(\gamma=0)= 2\hat{\phi}_p(k_c)=2P_1,\\
    \lambda(qk_c, \gamma=0)-2\lambda(k_c, \gamma=0)&=&r[2P_1-P_q+\beta(D_q-2D_1+1)]=r\Delta_q,\\
    c_0&=&c(\gamma=0)=\frac{P_1}{r}\left( \frac{P_1+P_2}{\Delta_2}- 2\frac{1+P_1}{1+ 2 \mu_0/r} \right)= \frac{P_1}{r}\mathcal{C},\\
    \nu_0&=&\nu(\gamma=0),\\
    c_0-\nu_0 &=& \frac{P_1}{r} \left( \frac{P_1+P_2}{\Delta_2}-2\frac{P_1+ P_{\sqrt{3}}}{\Delta_{\sqrt{3}}}\right)=\frac{P_1}{r}\mathcal{B}.
\end{eqnarray*}

\subsection{Instability of gap patterns}
The linear stability analysis for the pairwise nonlocal FKPP equation yields the eigenvalues
$$\lambda(k)= b(\hat{\phi}_D -1) - r \hat{\phi}_p.$$
Because $\phi_D$ is a probability density and radially symmetric, its Fourier transform is always less than one for $\mathbf{k}\neq 0$. This is particularly true for the Gaussian dispersal considered. Then, $(\hat{\phi}_D -1)<0$ always. A Turing instability requires that, for $k_c\neq0$, $\lambda(k_c)\geq0$. For this to hold, it must happen that
$$P_1 \leq \beta(D_1-1) \leq0.$$
Recalling the eigenvalue of the global phase for gap patterns:
$$\lambda_\Phi^{\text{gaps}}(\gamma)= 3 \rho_+ \cos (\pi) \kappa,$$
and $\kappa_0=2P_1<0$, we have 
$$\lambda_\Phi^{\text{gaps}}(\gamma=0) \geq,0$$
and thus gaps are unstable, or marginal at best.

\subsection{Instability of stripe patterns}
In this section, we reduce the conditions for stripe-pattern instability to three inequalities on the competition kernel. These conditions hold for super-Gaussian competition kernels and Gaussian dispersal kernels.
We first recall the stripe-pattern nontrivial eigenvalues
$$\lambda_\pm^{\text{stripes}}= \mu(1- \frac{\nu}{c}) \pm \kappa \sqrt{\frac{\mu}{c}}.$$
As $\kappa_0<0$, the greatest of the two corresponds to the minus sign. Evaluating at $\gamma=0$ it can be written as
$$\lambda_-^{\text{stripes}}(\gamma=0)= -\sqrt{\frac{\mu_0}{c_0}} P_1\left(2 - \mathcal{B}\sqrt{ \frac{\mu_0}{r} \frac{1}{P_1 \mathcal{C} }} \right).$$
$\sqrt{\mu_0/c_0}$ is just the amplitude of the stripe pattern predicted at leading order in the amplitude, and thus always positive. With $P_1<0$, the stripe pattern can be stable if the last square root term becomes large enough while $\mathcal{B}>0$.

\textbf{The general condition}. The condition for a stripe pattern to be unstable in the traditional nonlocal FKPP equation, for any kernels and parameters, is that
$$\mathcal{B}\sqrt{ \frac{\mu_0}{r} \frac{1}{P_1 \mathcal{C} }} <2.$$
For $\mathcal{B}<0$ the condition is trivially satisfied, so we restrict our analysis to $\mathcal{B}>0$. The Turing instability condition implies that $\mu_0/r < -P_1$, and thus
$$ \mathcal{B}\sqrt{ \frac{\mu_0}{r} \frac{1}{P_1 \mathcal{C} }} < \mathcal{B} \sqrt{\frac{1}{-\mathcal{C}}}.$$
Therefore, a sufficient condition for them being unstable is that
$$\mathcal{B} \sqrt{\frac{1}{-\mathcal{C}}}<2.$$

\textbf{A minimal set of conditions}. Consider a Gaussian dispersal kernel and the following set of conditions
\begin{eqnarray}
    P_1>-\frac{1}{6},\label{P_1}\\
    P_2>\frac{P_1}{2},\label{P_2}\\
    P_{\sqrt{3}}<-P_1.\label{P_3}
\end{eqnarray}
Then, the previous terms can be bounded. Recall that 
$$\Delta_q = 2P_1-P_q+\beta(D_q-2D_1+1).$$
Using the fact that $\mu_0/r= \beta(D_1-1)-P_1>0$, with \eqref{P_1} yields $$D_1\geq 1+ \frac{P_1}{\beta}\geq \frac{5}{6}.$$
Using the Gaussian dispersal kernel, $D_2-2D_1+1 = D_1^4-2D_1+1=(D_1-1)(D_1^3+D_1^2+D_1-1)$. It follows that $D_1-1 <0$, and that $(D_1^3+D_1^2+D_1-1)=D_1(D_1^2+D_1+1)-1>\frac{5}{6}(\frac{25}{36}+\frac{5}{6}+1) -1>0$. Therefore $D_2-2D_1+1<0$ and similarly $D_{\sqrt{3}}-2D_1+1<0$. We use these relations to bound
$$\Delta_2<2P_1-P_2,$$
and
$$\Delta_{\sqrt{3}}<2P_1- P_{\sqrt{3}}.$$
Moreover, the reduction to the resonant triad requires that all non-critical eigenvalues remain negative. Thus, $\lambda(qk_c)<0$, and therefore $\Delta_q<0$. 
Using \eqref{P_3}, we can bound $$2\frac{P_1 + P_{\sqrt{3}} }{\Delta_{\sqrt{3}}} > 0,$$ yielding
$$\mathcal{B}< \frac{P_1+P_2}{\Delta_2}.$$
As $\mathcal{B}>0$, it follows that $P_1+P_2<0$, and we can bound
$$\mathcal{B}<\frac{P_1+P_2}{2P_1-P_2}<1,$$
which follows from \eqref{P_2}.

Because $\mathcal{B}<1$, we can further bound

$$ \mathcal{B} \sqrt{\frac{1}{-\mathcal{C}}}< \sqrt{\frac{1}{-\mathcal{C}}}.$$

The instability condition now depends on $\mathcal{C}.$ Recalling

$$\mathcal{C}= \frac{P_1+P_2}{\Delta_2}- 2\frac{1+P_1}{1+ 2 \mu_0/r},$$
$1+P_1>0,$ and $\mu_0/r<-P_1,$ we obtain
$$ 2\frac{1+P_1}{1+ 2 \mu_0/r} > 2\frac{1+P_1}{1-2P_1}.$$
Combining it with the previous bound, 
$$\mathcal{B}< \frac{P_1+P_2}{\Delta_2}<\frac{P_1+P_2}{2P_1-P_2}<1,$$
it follows that
$$ \mathcal{C}< 1- 2\frac{1+P_1}{1-2P_1} = - \frac{1+4P_1}{1-2P_1}< -\frac{1}{4},$$
which is obtained by using the condition \eqref{P_1}. Therefore, we have demonstrated that the conditions \eqref{P_1}-\eqref{P_3} lead to
$$\mathcal{B}\sqrt{ \frac{\mu_0}{r} \frac{1}{P_1 \mathcal{C} }}<2,$$
constituting the sufficient conditions that make stripe patterns unstable under Gaussian dispersal in the traditional nonlocal FKPP equation.

\begin{figure}[t!]
	\centering
	\includegraphics[]{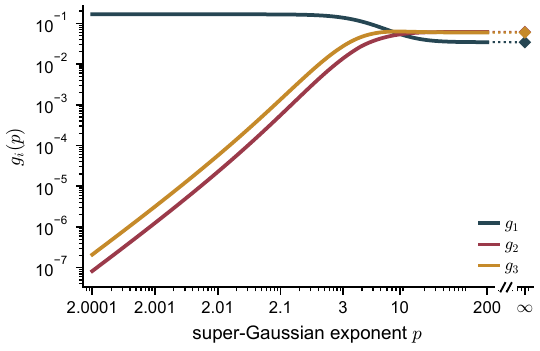}
	\caption{Numerically verified bound for conditions \eqref{P_1}-\eqref{P_3}.}
	\label{FS2}
\end{figure}

\textbf{The super-Gaussian competition kernel}. The conditions \eqref{P_1}-\eqref{P_3} hold for an arbitrary super-Gaussian competition kernel under Gaussian dispersal. We prove this by considering that any critical wavenumber must satisfy

$$\hat{\phi}_p(k_c)<0,$$
$$\beta \partial_k\hat{\phi}_D|_{k_c}= \partial_k \hat{\phi}_p|_{k_c}.$$
Because the dispersal is Gaussian, $\partial_k \hat{\phi}_D<0$, and $\partial_k \hat{\phi}_p|_{k_c}<0$ follows. Noting that
$$\partial_k \hat{\phi}_D = -s^2k e^{-s^2k^2/2},$$
and using $xe^{-x}< 1-e^{-x}$ on $\beta(\hat{\phi}_D-1)-\hat{\phi}_p>0$ leads to
$$ 0< -\beta(1-\hat{\phi}_D)-\hat{\phi}_p< -\beta \frac{s^2 k^2}{2}e^{-s^2k^2/2} - \hat{\phi}_p = \frac{k}{2} \partial_k \hat{\phi}_p - \hat{\phi}_p.$$
Therefore, using nondimensional spatial units, $z= r_c k$, $r_c\partial_z=\partial_k$, any selected wavenumber $z_c$ is bounded by the conditions
\begin{eqnarray*}
    \hat{\phi}_p |_{z_c}<0, \\
    \partial_z \hat{\phi}_p|_{z_c}<0,\\
    \left(\hat{\phi}_p - \frac{z}{2}\partial_z \hat{\phi}_p\right)\bigg|_{z_c}<0.
\end{eqnarray*}
We call the group of all $z_c$ satisfying the above conditions $Z_c$. Then, proving that the super-Gaussian kernel satisfies the conditions \eqref{P_1}-\eqref{P_3} reduces to prove that they hold for any $z \in Z_c$. Then, we define
\begin{eqnarray*}
    G_1(p, z)&=& \hat{\phi}_p(z) + \frac{1}{6},\\
p    G_2(p, z)&=& \hat{\phi}_p(2z)- \frac{1}{2}\hat{\phi}_p(z) ,\\
    G_3(p, z)&=& -\hat{\phi}_p(\sqrt{3}z)- \hat{\phi}_p(z),
\end{eqnarray*}
and
$$g_j(p)= \min_{z\in Z_c} G_j(p, z).$$
If the functions $g_j>0$ for all $p>2$, then the conditions \eqref{P_1}-\eqref{P_3} are always fulfilled. These functions are plotted in Fig. \ref{FS2} for $p>2$ up to $200$, with the values of the top-hat limit, $p\to\infty$ shown as well.

\end{document}